\documentclass[11pt,a4paper]{article}
\usepackage{geometry}
\usepackage{amsthm}
\usepackage{authblk}             	               
\usepackage[parfill]{parskip}
\usepackage{pgfplots}  
\usepackage{graphics, color}
\usepackage{floatflt}				
\usepackage[T1]{fontenc}
\usepackage[english]{babel}
\usepackage[latin1]{inputenc}			
\usepackage{amssymb}

\begin{document}

\title{The astronomical garden of \textit{Venus and Mars - NG915}: 
\\ the pivotal role of Astronomy in dating and deciphering Botticelli's masterpiece}

\author{Mariateresa Crosta}
\affil{Istituto Nazionale di Astrofisica (INAF- OATo), \\ Via Osservatorio 20, Pino Torinese -10025, TO,  Italy \\e-mail: crosta@oato.inaf.it }

\date{}

\maketitle

\begin{abstract}

This essay demonstrates the key role of Astronomy in the Botticelli {\it Venus and Mars-NG915} painting, to date only very partially understood.  Worthwhile coincidences among the principles of the Ficinian philosophy, the historical characters involved and the compositional elements of the painting, show how the astronomical knowledge of that time strongly influenced this masterpiece. First, Astronomy provides its precise dating since the artist used the astronomical ephemerides of his time, albeit preserving a mythological meaning, and a clue for Botticelli's signature.  Second, it allows the correlation among Botticelli's creative intention, the historical facts and the astronomical phenomena such as the heliacal rising of the planet Venus in conjunction with the Aquarius constellation dating back to the earliest representations of Venus in Mesopotamian culture. This work not only bears a significant value for the history of science and art, but, in the current era of three-dimensional mapping of billion stars about to be delivered by Gaia, states the role of astronomical heritage in Western culture. Finally, following the same method, a precise astronomical dating for the famous {\it Primavera} painting is suggested.

\end{abstract}


\section*{Introduction}

Since its acquisition by London's National Gallery on June 1874,  the painting  {\it Venus and Mars}  by Botticelli, cataloged as {\it NG915}, has remained a mystery to be interpreted \cite{paoli}\footnote{For an exhaustive list of all the interpretations of NG915, please refer to the recent book by Marco Paoli {\it Venere Marte, Parodia di un adulterio nella Firenze di Lorenzo il Magnifico}.}.
In the present study the association of an astronomical configuration has been determinant for reading this masterpiece, and for its accurate dating.  A  search on the astronomical charts of the late '400 has widely supported the initial intuition. 

\begin{figure}
\centering
\begin{minipage}{1.0\textwidth}
   \includegraphics[width=\textwidth]{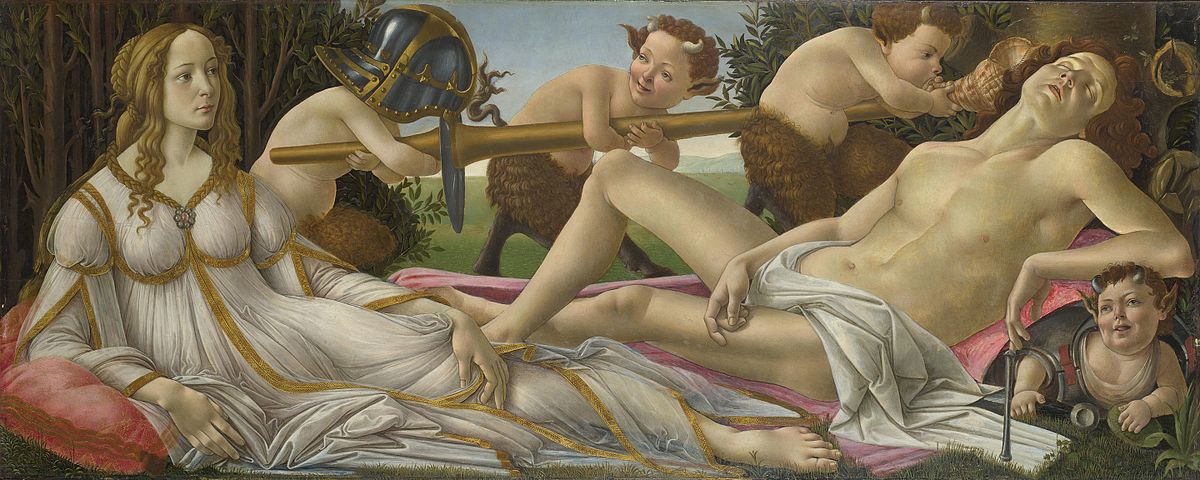}
 \end{minipage}
 \hfill
\begin{minipage}{0.45\textwidth}
\centering
\includegraphics[width=\textwidth]{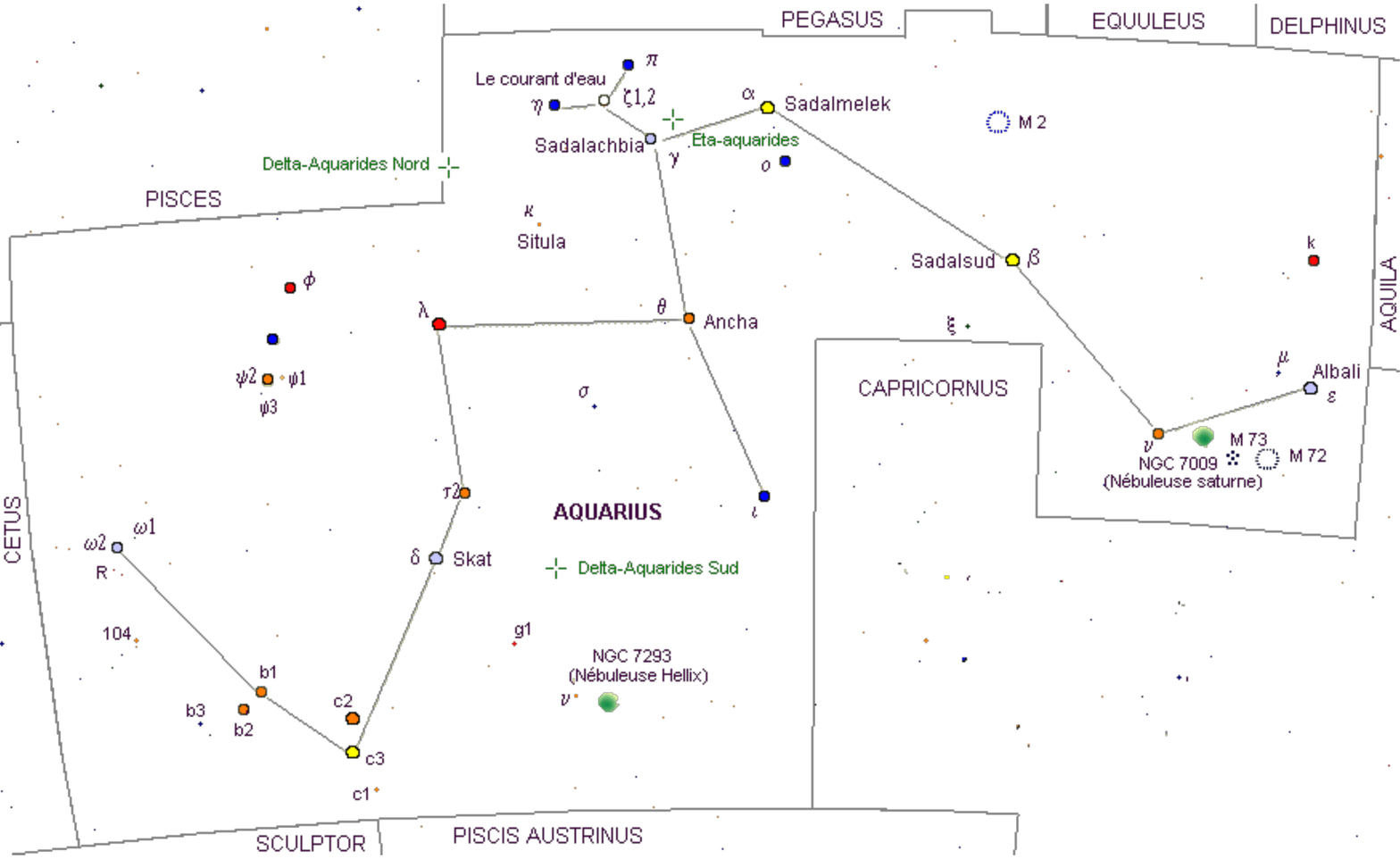}
\end{minipage}
\hfill
\begin{minipage}{0.45\textwidth}
\centering
\includegraphics[width=\textwidth]{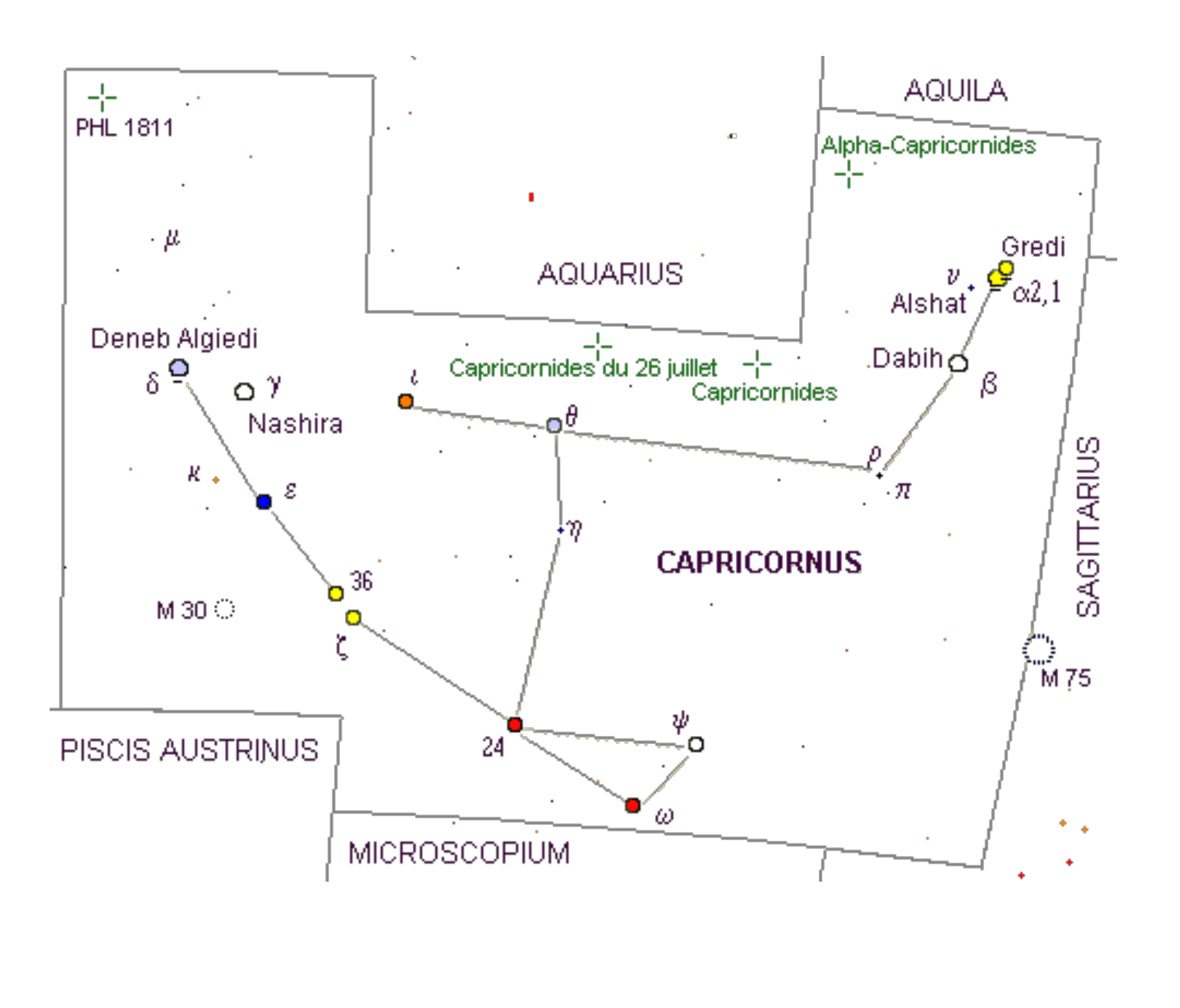}
\end{minipage}
\hfill
\begin{minipage}{0.9\textwidth}
\includegraphics[width=\textwidth]{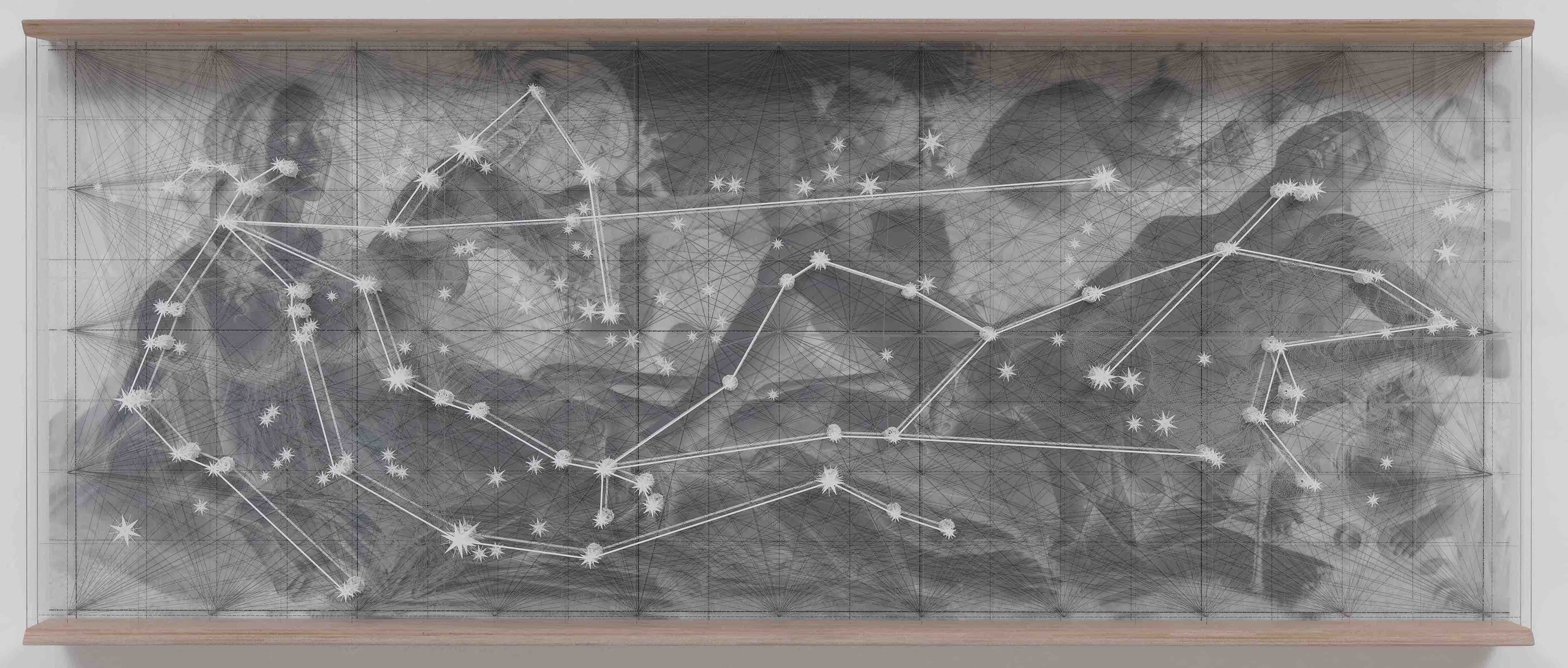}
\end{minipage}
\caption{\small{\label{Fig1} Top:  NG915 - {\it Venus and Mars}, Botticelli, London's National Gallery. Center: astronomical charts of Acquarius and Capricon.  Bottom:  {\it Study for venus and mars garden project 148*/2017} by Michele Guido (courtesy by Lia Rumma Gallery, photo by Antonio Maniscalco), NG915 astral free adaptation (from {\it Uranometria} ) inspired by the interaction of the artist with astrophysicist Mariateresa Crosta for the tenth edition of {\it Meteoriti in Giardino}  (organized by Merz Foundation).}}
\end{figure}

Figure \ref{Fig1} presents Botticelli's painting associated to a picture showing a relationship between the representation of  {\it Venus and Mars} and some possible asterisms. The link arose in the context of the space mission Gaia (European Space Agency, ESA \cite{gaia}), currently in orbit at 1.5 million km from Earth, whose goal is a high-precision three-dimensional map of Galaxy. Indeed, Gaia will generate the most important and most numerous (about 2 billion objects) astronomical cartography ever realized by humanity. 

Given that one can trace as many asterisms as the number of visible stars, it was necessary to find a clue for the proper interpretation of NG915. Thanks to the similarity of the dress worn by Venus in {\it NG915} with that one in the famous {\it Primavera}, the first step was checking the astral situation of the Spring Equinox at the latitude of Florence around the presumable years dating the painting.

Among the possible constellations, Aquarius and Capricorn turned out to be the best candidates. In the northern hemisphere they are visible in the summer and fall night sky, but the heliacal rising of both happens toward the equinox of March. Let us remind that heliacal rising means the rising of a star just before dawn, consequently its visibility in the morning until the sunlight diffuses completely. 
Moreover in this circumstance Aquarius and Capricorn appear in some cases in conjunction with planets Venus and Mars respectively.

\section*{Premise: the astronomical atlases at Botticelli's time} 
The need to position the stars and the resulting study is the most ancient branch of Astronomy named Astrometry. Comparing the positions of stellar objects at different epochs allows to determine their distance and their proper motion. The measurements to position the stars were carried out over the centuries with increasingly sophisticated instrumentation and today also from space, as in the case of the Gaia mission.
Only since 1930 the International Astronomical Union has adopted criteria to compare the various celestial charts and established the current 88 constellations. Until then the shape and boundaries of the constellations, as well as their names, although passed on, were still susceptible of free interpretation by astronomers.
The definitive IAU constellations include all those described by Ptolemy in the famous {\it Almagesto}, with the addition of those of {\it Uranometria} by Johann Bayer, the first atlas that covered the entire celestial sky, i.e. including also the southern hemisphere, published in 1603.
Prior to Bayer's {\it Uranometria}, the fundamental text was Ptolemy's Almagest, representing the culmination of the scientific production of Greek astronomers and philosophers such as Eratosthenes, Hipparchus and Ptolemy himself. The {\it Almagesto} constituted the scientific context throughout the Middle Ages and the Renaissance and was only updated in the {\it Liber locis stellarum fixarum} by Abd-al-Rahman al-Sufi in 964 \footnote{A part from a very limited number of Arabic globes, there remains a representation of the sky of Hipparchus and Ptolemy thanks to the celestial globe on the shoulder of {\it Atlante Farnese} in the National Archaeological Museum of Naples.}.

For the purpose of this study, it is worth mentioning that before {\it Uranometria}, {\it De le stelle fisse}, published in 1543 by Alessandro Piccolomini, was the first modern celestial atlas and the first to assign Latin letters to the stars according to their luminosity. The maps contained in that work include all the Ptolemaic constellations (except one) and show the stars without the corresponding mythological figures; so we infer that, in any case, before that date, mythology was part of the interpretation of the sky.

As matter of fact until the 15th century the didactic poem in hexameters {\it Phaenomena} by Arato had been in circulation for a long time. Such a work included the millennial celestial knowledge received and elaborated by Greeks, and transmitted to Lucrezio, Virgilio, Cicero, Ovid, Germanicus, and Avienio.\footnote {One can add also the translations of Germanicus between I sec. B.C. and I sec. d.C., Avienio in IV sec. and those of an anonymous monk of the French monastery of Corbie in the eighteenth century, {\it Aratus Latinus Primitivo}. Some of these translations (those of Cicero, Germanic and Aratus Latinus) have been handed down to us in manuscripts with illustrations and in some cases they present traces of their precursor models.}
 Arato's translations includes also two short works of the Early Middle Ages, the {\it De signis caeli} and {\it De ordine ac positione stellarum in signis}, and Igino's {\it De Astronomia} (C. Julius Hyginus, 64 BC-17 AD) manuscripts. Igino resumes materials dating back to Eratosthenes of Cyrene, who in the III sec. B.C. wrote the astronomical treatise, {\it Catasterismi}, which means ``transformation into stars''. The second book by Igino describes the mythological stories for each constellation that constitute the basis on which the constellation itself have been formed, and its process of collocation in the sky (i.e. {\it catasterismo}).

Igino's {\it De Astronomia} has been passed on through numerous independent medieval manuscripts and print works published between the fifteenth and seventeenth centuries, containing graphic representations of mythological characters not always philologically consistent with the text.
In such works the aesthetic and astrological needs prevail, the philological and literary interpretation is preferred to such a degree that the positioning of the stars is conditioned to coincide with the drawing of an anatomical detail of the mythological figure to which they belong. 

Al-Ma'mun \footnote{Son of Harun al-Rashid, who set up a personal library, called Bayt al-Hikma, "The House of Science," which Al-Ma'mun enlarged to create the richest library of the whole Islamic world. } explains the reason for naming constellations rather than single stars: \\ << There are many stars in everywhere and many of them are identical in size and brightness in their travels. For this reason it seemed reasonable to group the stars together, so that, arranged with one another, they represented figures, so the stars became nominated >> \footnote{Author's translation from italian.}.
But even when the stars are singly named, the meaning of the translation of their Greek, Latin, and Arabic traditional name almost always identifies the anatomical position or a quality which the star occupies in the figure. And this tradition will last throughout the Middle Ages and the Renaissance, and will be interrupted right in Bayer's atlas by  D\"urer.

The numerous editions of Igino's tales constitued the {\it Poeticon Astronomicon}. The first print publication was edited in Venice in 1482, on behalf of Erhard Ratdolt.

As briefly outlined\footnote{For details, references \cite{stoppa,stoppaatlanti}.} this was essentially the corpus of the astronomical knowledge to which Sandro Botticelli could have been exposed at the time he lived and worked in Florence under the influence of the Medici family.  The interest towards the classical culture was renewed thanks to the translations of Plotin's treatises and Plato's dialogues by Marsilio Ficino and the creation of the Neoplatonic Academy, whose principles were fully expressed through multiple readings in Botticelli's works. Marsilio Ficino wrote:<<According to the most ancient followers of Plato, the Soul of the World has built beside the stars figures and portions of figures that are themselves figures of a certain type, also conferring certain properties on each of them. In the stars - in their figures, parts and properties - are contained all sorts of things that are in the lower world and their properties>>\footnote{Author's translation from: <<Secondo infatti i pi\`u antichi platonici dalle sue ragioni l'Anima del Mondo ha costruito accanto alle stelle figure e parti di figure tali che sono esse stesse figure di un certo tipo conferendo anche determinate propriet\`a a ciascuna di esse. Inoltre nelle stelle - nelle loro figure, parti e propriet\`a - sono contenute tutte le specie di cose che si trovano nel mondo inferiore e le loro propriet\`a>>.}.

\section*{Dating NG915 according to the motion of Mars and Venus from 1480 to 1488} 
\begin{figure}
\centering
\includegraphics[width=15.0cm]{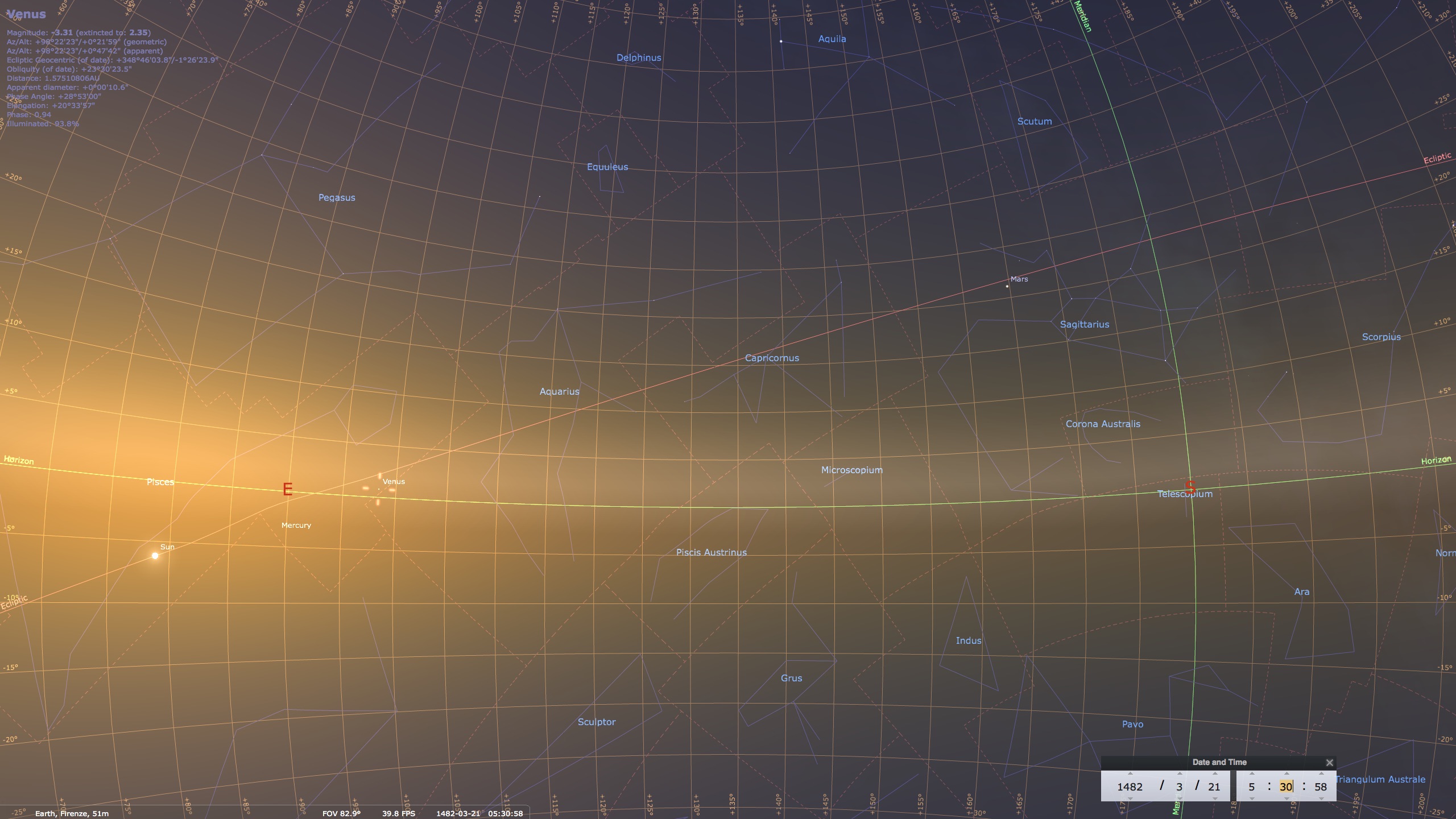}
\caption{\label{Fig2}Aquarium and Capricorn around dawn around the Spring Equinox at the latitudes of Florence in 1482.}
\end{figure}
\begin{figure}
   \centering
\begin{minipage}{0.95\textwidth}
   \centering
   \includegraphics[width=\textwidth]{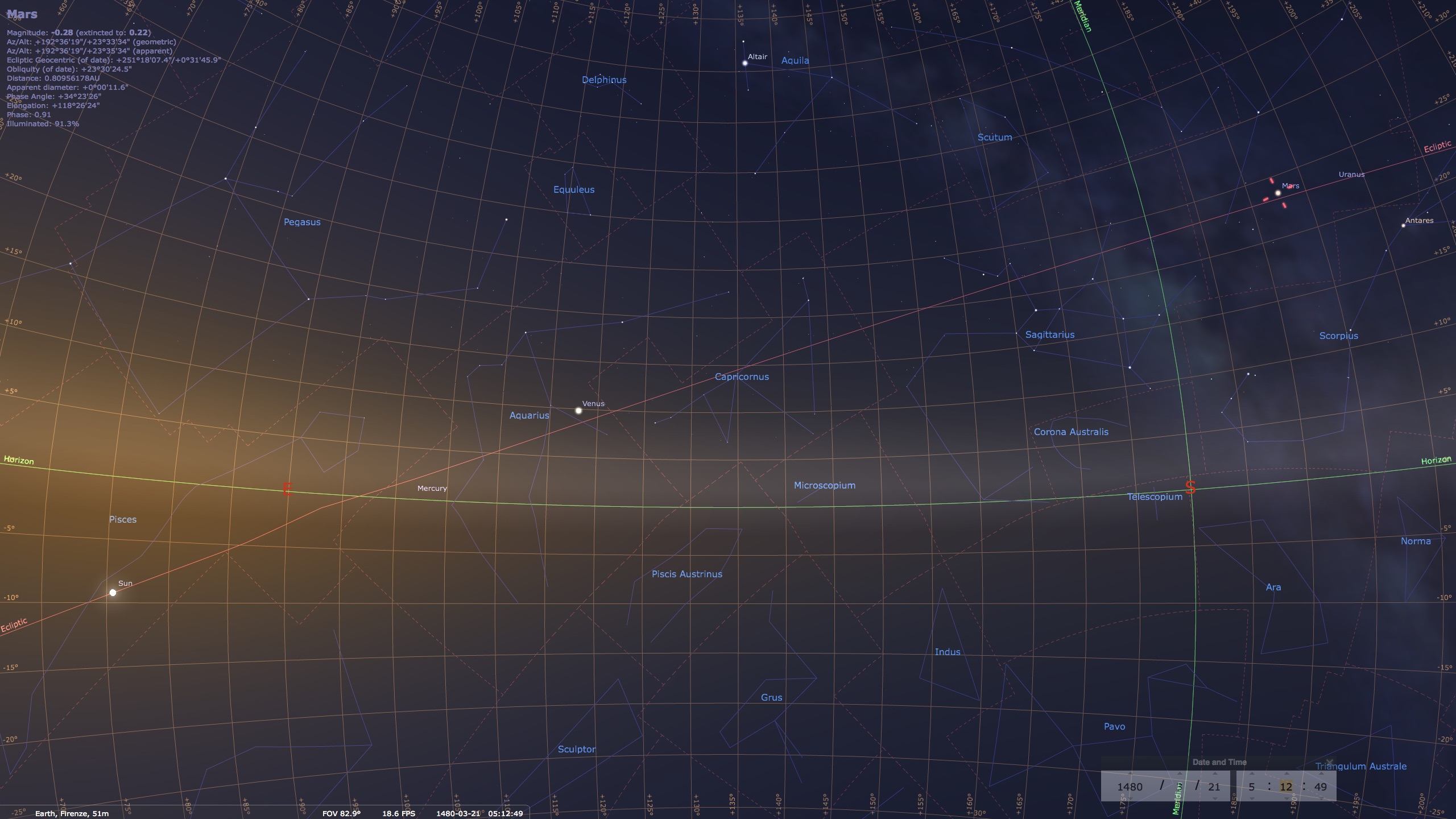} 
\end{minipage}
\hfill
\begin{minipage}{0.95\textwidth}
   \centering
   \includegraphics[width=\textwidth]{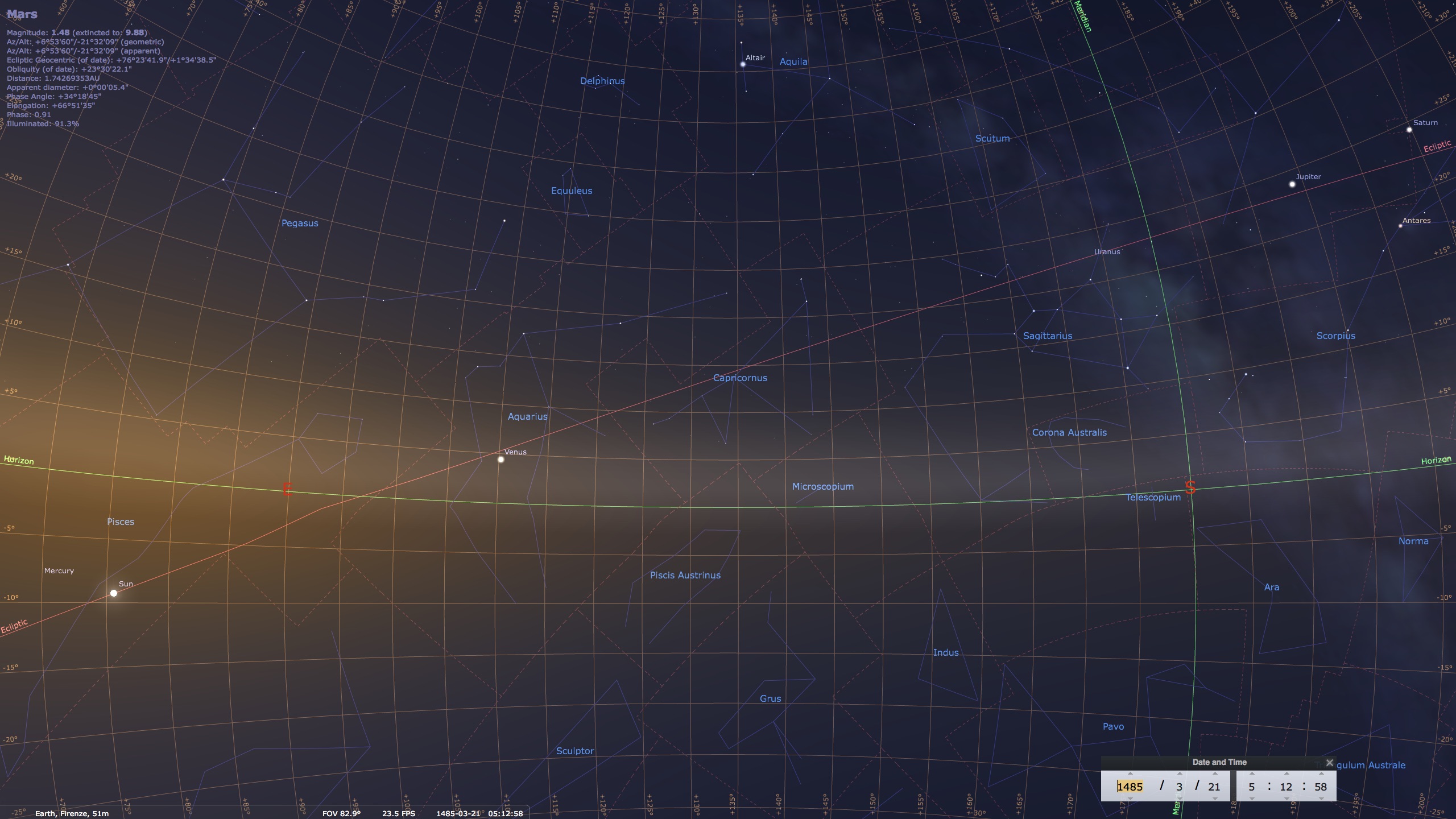}
  \end{minipage}
\begin{minipage}{0.95\textwidth}
\centering
\includegraphics[width=\textwidth]{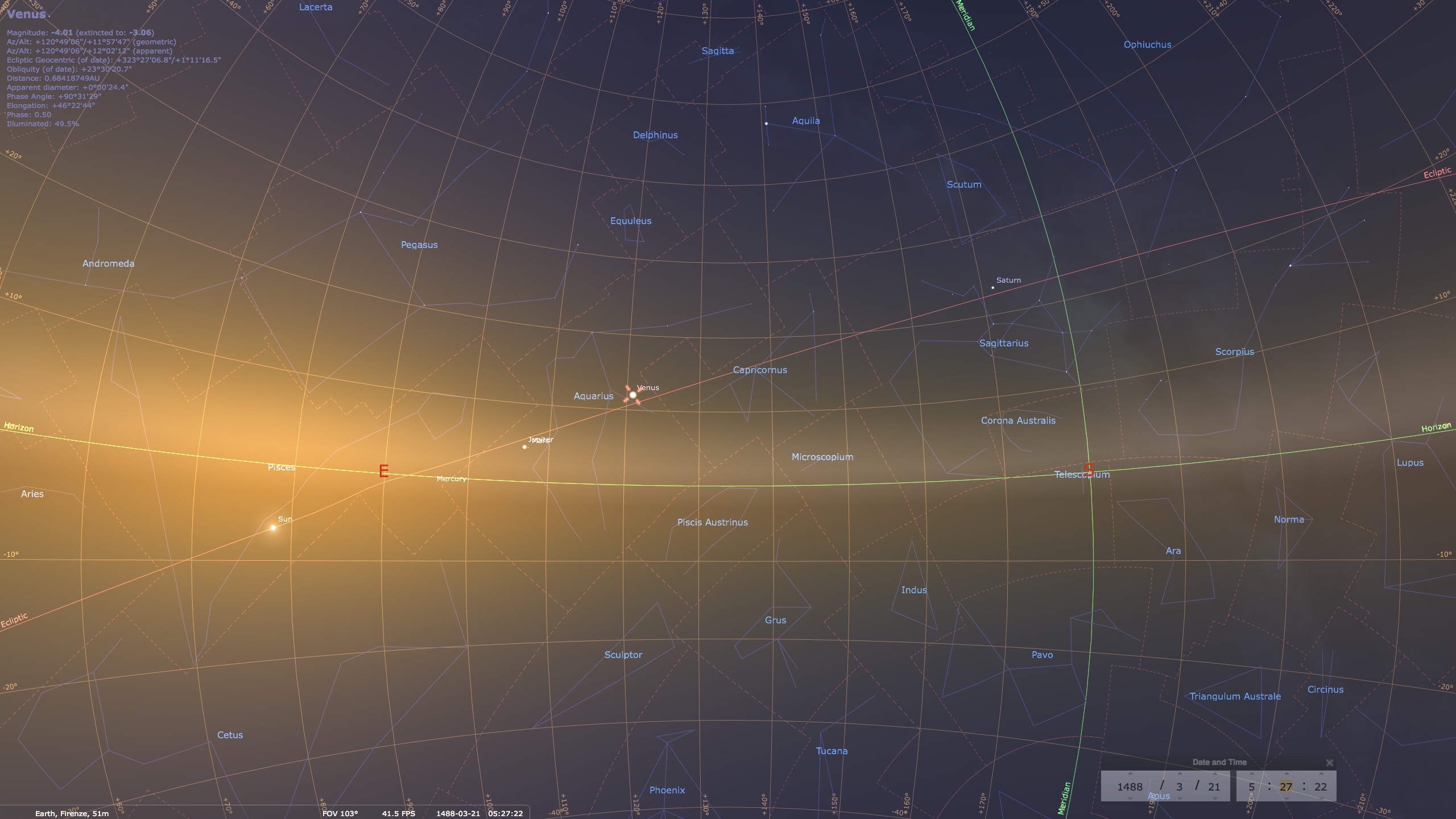}
\end{minipage}
\caption{\small{\label{Fig3}Heliacal rising of Aquarius and Capricorn around the spring equinox in Florence in 1480,1485, and 1488. Note the alignment between Mars and Jupiter in 1488.}}
\end{figure}
\begin{figure}
   \centering
\begin{minipage}{0.95\textwidth}
   \centering
   \includegraphics[width=\textwidth]{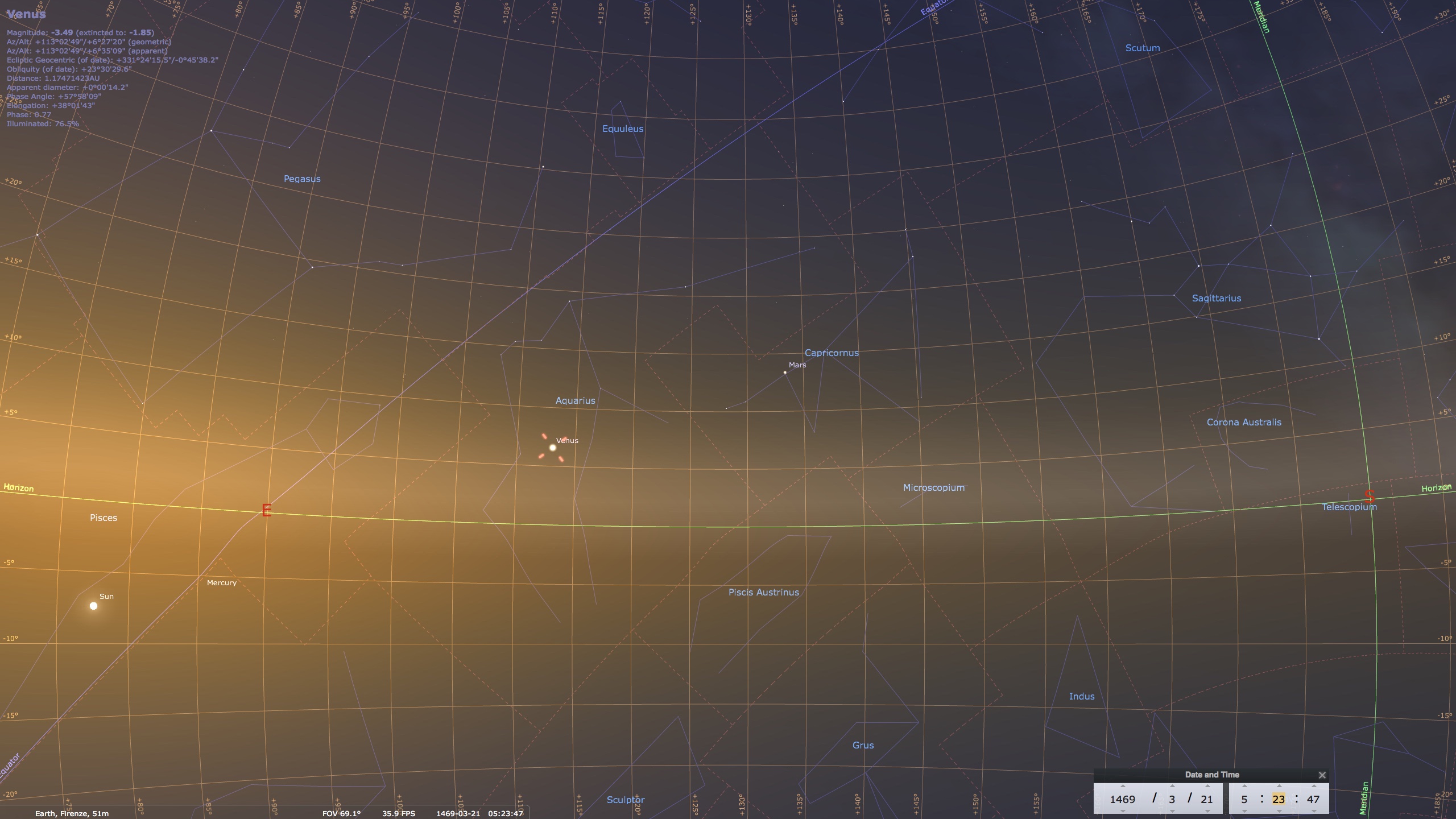}    
   \end{minipage}
\hfill   
\begin{minipage}{0.95\textwidth}
   \centering
   \includegraphics[width=\textwidth]{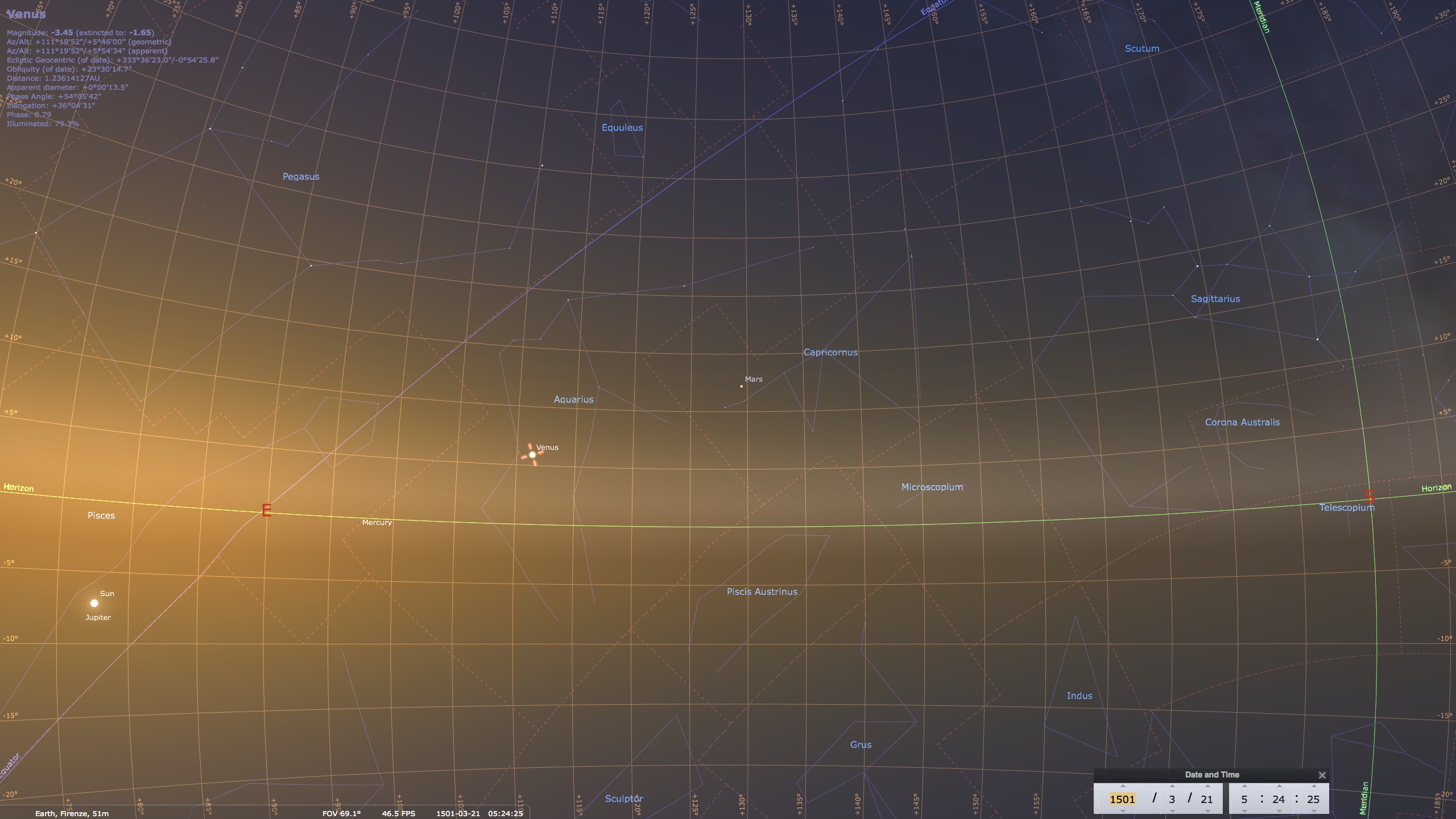}    
   \end{minipage}
    \caption{\label{Fig4}The ascending Aquarius and Capricorn at dawn around the spring equinox at the latitudes of Florence in 1469 and 1501; note Venus and Mars in conjunction with the constellations.}
\end{figure}

Around the spring equinox of 1482, at the latitudes of Florence, Venus is Eastward of Aquarius, in heliacal rising, 68 degrees far from the planet Mars which is West of Capricorn in the constellation of Sagittarius (figure \ref{Fig2})\footnote{It is worth mentioning that at the time of Botticelli the Gregorian calendar had not yet entered into force, so in theory the date of the equinox should be backdated by 10 days, but astronomical software uses the Julian date converter.}.  
This coincidence offers a unique dating of the painting against the uncertainties that date it in an interval ranging from 1480 to 1488. In fact at the spring equinox of 1480 Mars is in the Ofiuco constellation, in 1485 it does not appear and in 1488 it  is in conjunction with Jupiter at Eastward of Aquarius and Venus on West (figure \ref{Fig3}). In the remaining dates, nothing relevant  for the purpose of this study is observed.
 A part form this time interval, the exact simultaneous presence of Venus, in conjunction with Aquarius, and Mars, in conjunction with Capricorn, occurs only in 1469 and in 1501 (figure \ref{Fig4}), dates not consistent with the Botticelli biography. 
Mars generally looks orange or reddish, with brightness quite variable throughout its orbit. The apparent magnitude of Mars passes from +1.8 mag (visual band) at conjunction to the much brighter -2.9 at perihelion opposition, phenomenon that occurs every two years and makes the planet difficult to observe. The ancients were already aware of its retrograde motion.

Venus, on the other hand, is yellowish-whitish and makes a revolution along an almost circular orbit in 224.7 Earth days. Known since ancient times as the brightest natural object in the night sky, after the Moon, it reaches an apparent maximum visual magnitude of -4.4.

Being an inner planet, it is only visible shortly before dawn or shortly after sunset - hence the respective denominations {\it Lucifero} and {\it Vespero} - and for a few hours near the Sun, then moves alternately east and west of the Sun. In fact, the favorable periods for observing the planet are those where elongation - the angular distance between a planet and the Sun- reaches the highest values of 47 degrees East or 47 degrees West: in the first case, the planet appears immediately after sunset, in the second just before dawn.

Let us also remind that apart from the Sun, the Moon and rarely Jupiter, Venus is the only celestial body visible to the naked eye during daylight, provided that the sky is not cloudy and its elongation from the Sun is not at the minimum.

\section*{Aquarius constellation in NG915, the connection with the planet Venus and Botticelli's signature } 

Aquarius is a winter constellation, considered in ancient times as a rain carrier. Its connection with water goes back to Babylonian astronomy, so nearby it on the sky we find the Fish and other aquatic constellations, such as Ceto and Capricorn, located in the  so-called  `` Celestial Waters ''  region or ``The Sea'' \footnote{Here and after for the astronomical details refer to \cite{certissimasigna}.}.

In the {\it Catasterismi} this constellation is often represented by Ganimede, the young Trojan that Zeus rapes for his beauty and transports to heavens to serve the gods as a cupbearer, as Ovid told. Zeus then donates Ganimede the immortality by transforming him into the constellation of Aquarius. Neoplatonism provides a mystical representation of Ganymede's abduction, meaning the abduction of the soul to God, whereas in ancient Greece and Rome the myth became very popular and considered a divine endorsement of homosexuality.
 
Another alternative association is Deucalion, who wandered with its arch on flood waters for nine days and nine nights, but Igino, citing Eubulo, still offers another identification of the constellation with Cecrope, the first mythical king of Athens: since he reigned in times when the wine was not yet invented, he is depicted while offering a sacrifice to the gods with water.

Infact the Aquarius constellation is generally represented by the figure of a man with an arm open in the direction of Capricorn holding the flap of a mantle or a crossbar while the other arm, whose hand is almost in contact with Pegasus, holds an amphora from where the {\it Fluvius Aquarii} flows up to the Southern Fish. Dante described his appearance in the night sky as the period when winter slowly moves towards spring.

For the present study is also interesting is what Aquarius meant for the Babylonians. Sumerian called it the <<Great Man>> (GU.LA) and it was identified with the god Ea, the Lord of the source and depicted while holding a jar in his hand from which two water streams spurt. However Ea was also the symbol of Capricorn.
\begin{figure}
 \centering
\includegraphics[width=14.5cm]{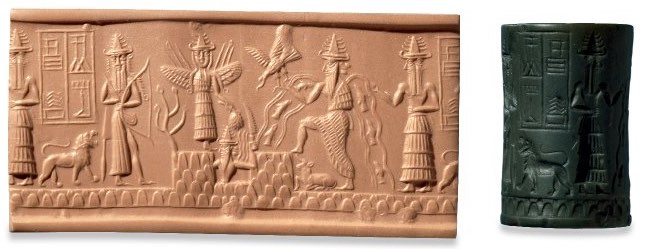} 
\caption{\label{Fig6}The Adda seal, British Museum, Museum number 89115}
\end{figure}
\hfill
\begin{figure}
\begin{minipage}{0.35\textwidth}
   \centering
   \includegraphics[width=\textwidth]{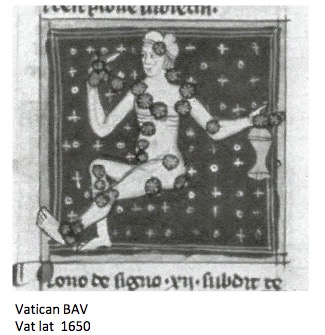} 
 \end{minipage}
\hfill
\begin{minipage}{0.28\textwidth}
   \centering
   \includegraphics[width=\textwidth]{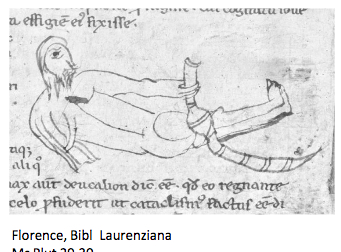}  
   \end{minipage}
\hfill
\begin{minipage}{0.35\textwidth}
\centering
\includegraphics[width=\textwidth]{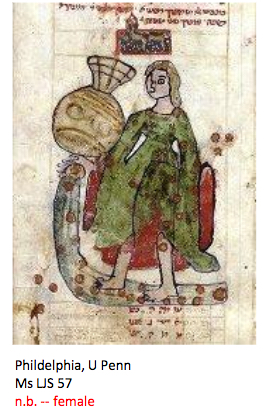}
\end{minipage}
\hfill
\begin{minipage}{0.3\textwidth}
\centering
\includegraphics[width=\textwidth]{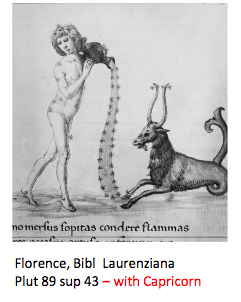}
\end{minipage}
\hfill
\begin{minipage}{0.28\textwidth}
   \centering
   \includegraphics[width=\textwidth]{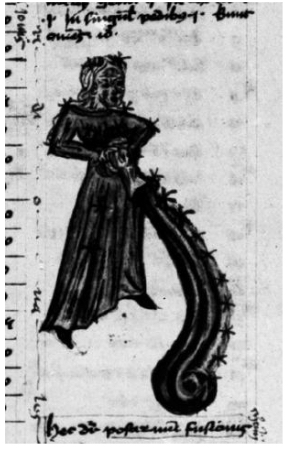}
   \end{minipage}
\hfill
\begin{minipage}{0.28\textwidth}
\centering
\includegraphics[width=\textwidth]{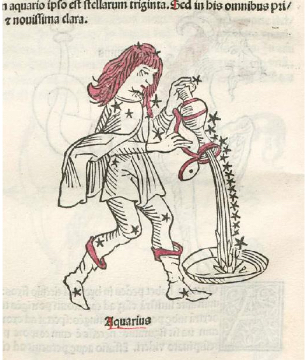}
\end{minipage}
 \caption{\label{Fig7} From upper left: the Aquarius constellation according to  Seneca, Tragedies (comm. Nicolas Trevet)\cite{lippncott}; {\it Hyginus Aquarius}, Florence, Bibl  Laurenziana, Ms. Plut 29.30\cite{lippncott}; {\it Astronomical Compendium} in Hebrew, Catalonia, c. 1361\cite{lippncott};  {\it Germanicus  Aratea,  Hyginus,  Astronomica}, Biblioteca Laurenziana, Plut. 89. sup 43 \cite{lippncott};  Acquarius full dressed, maybe female, `Ptolemaic' stellar tables, Vatican BAV Pal lat 1368 \cite{lippncott}; depiction of Aquarius in the publication of {\it Poeticon}, 1482.}
\end{figure}
\hfill
In addition, in the seal of Adda (figure \ref{Fig6}) Ea is the divinity of Earth and Life, who abides in abysmal waters, and is represented with two branches of water flowing to the ground \cite{cesta}. In the same seal is also depicted Ishtar, the goddess of love and war. Note the co-presence of Ishtar and Aquarius-Ea as if there was a connection with the well known greek myth that Venus is born from the sea foam. Indeed, the original representation of Babylonian divinity should be sought in the most ancient Sumerian culture - in the form of Enki - although the primordial reference could lead us to Vedic India, where the man's figure, Trita Aptya (a pre-Vedic god), was the officer who holds the vase in the <<ayoma>> ritual. Tritha became Triton in the cultures of the Mediterranean, a god with the lower limbs shaped as double tail of fish, who had in his hand a twisted shell whose sound was used to burst or calm storms. 
Going back to the possible representations of Aquarius, in 1482 (at the time of the publication of the {\it Poeticon Astronomicon}) such a constellation is depicted dressed with long and loose hair, where a folded right leg is evident. Below, in the figures \ref{Fig7}, a selection of the various pictures about Aquarius in the old atlases. For a complete list of images, refer to ``The Saxl Project''  \cite{lippncott} and Certissima Signa \cite{certissimasigna}. 
In particular, the reader should pay attention to the position of the right leg of the Aquarius figure, to the reclined ones and especially to the image in figure \ref{Fig7}, where  Capricorn is facing Aquarius like in NG915 instead to be in the opposite direction as in the traditional representation. Moreover in some of them Aquarius is a female character, with long hair, although is mostly male in others.

\paragraph{Simonetta Vespucci as Venus and Aquarius.} 
The woman in the foreground, Venus, as hypothesized by many critics, resembles Simonetta Vespucci and wears the wedding white dress with golden edges similar to that of Venus in the famous {\it Primavera} picture, so the first clue was looking for a reference to the spring equinox.
As anticipated, the constellations of Aquarius and Capricorn have heliacal rising at the spring equinox at the latitudes of Florence.
Historical sources do not provide Simonetta Vespucci's month and place of birth. She was born in 1453, perhaps in Portovenere - according to Poliziano ``in grembo a Venere'' -  and on January 28 \cite{govetti,allan} by the nobles Gaspare Cattaneo della Volta and Cattochia Spinola de Candia. If the day was correct, Simonetta would have been of the Aquarius sign, i.e. the Sun at the time of her birth was in Aquarius constellation. When she was sixteen, Simonetta was married to the young Florentine Marco Vespucci. The date of their marriage is supposed towards August 1469 \cite{biotreccani}. In other texts it is also indicated on the beginning of 1469 or April 1469 \cite{govetti}.
After the marriage the couple settled in Florence; their arrival coincided with the rise of Lorenzo dei Medici.

Simonetta died very young of phthisis on April 26, 1476, even Lorenzo sends his personal doctor as the last desperate attempt to save her life.
Dressed in white as a bride, Simonetta's coffin crossed Florence with her face and body uncovered, escorted to the burial place in the church of Ognissanti. On that occasion Pulci and Poliziano depicted her respectively as: 
<<Ma forse che ancor viva al mondo \`e quella poich\'e vista da noi fu dopo il fine in sul feretro posta assai pi\`u bella>> and <<Bellezza Immortale>> .

Lorenzo il Magnifico wrote for her a sonnet \footnote{ << O chiara stella che coi raggi tuoi / togli alle vicine stelle il lume, / perch\`e splendi assai pi\`u del tuo costume?/
Perch\`e  con Febo ancor contender vuoi?/
Forse i belli occhi, quali ha tolti a noi/
morte crudel, che ormai troppo presume,/
accolti hai in te: adorna del loro nume,/
il suo bel carro a Phebo chieder puoi./
O questo o nuova stella che tu sia,/
che di splendor novello adorni il cielo,/
chiamata essa ud\`i, nume, i voti nostri:/
leva dello splendor tuo tanto via,/
che agli occhi, che han d'eterno pianto zelo,/
sanza altra offension lieta ti mostri>>.}
apparently inspired by a very glittering star in the clear night - perhaps the planet Venus or a fireball? - so shiny that could only be Simonetta's luminous soul joining to an object of the firmament.
 \begin{figure}
   \centering
\begin{minipage}{1.0\textwidth}
   \centering
   \includegraphics[width=\textwidth]{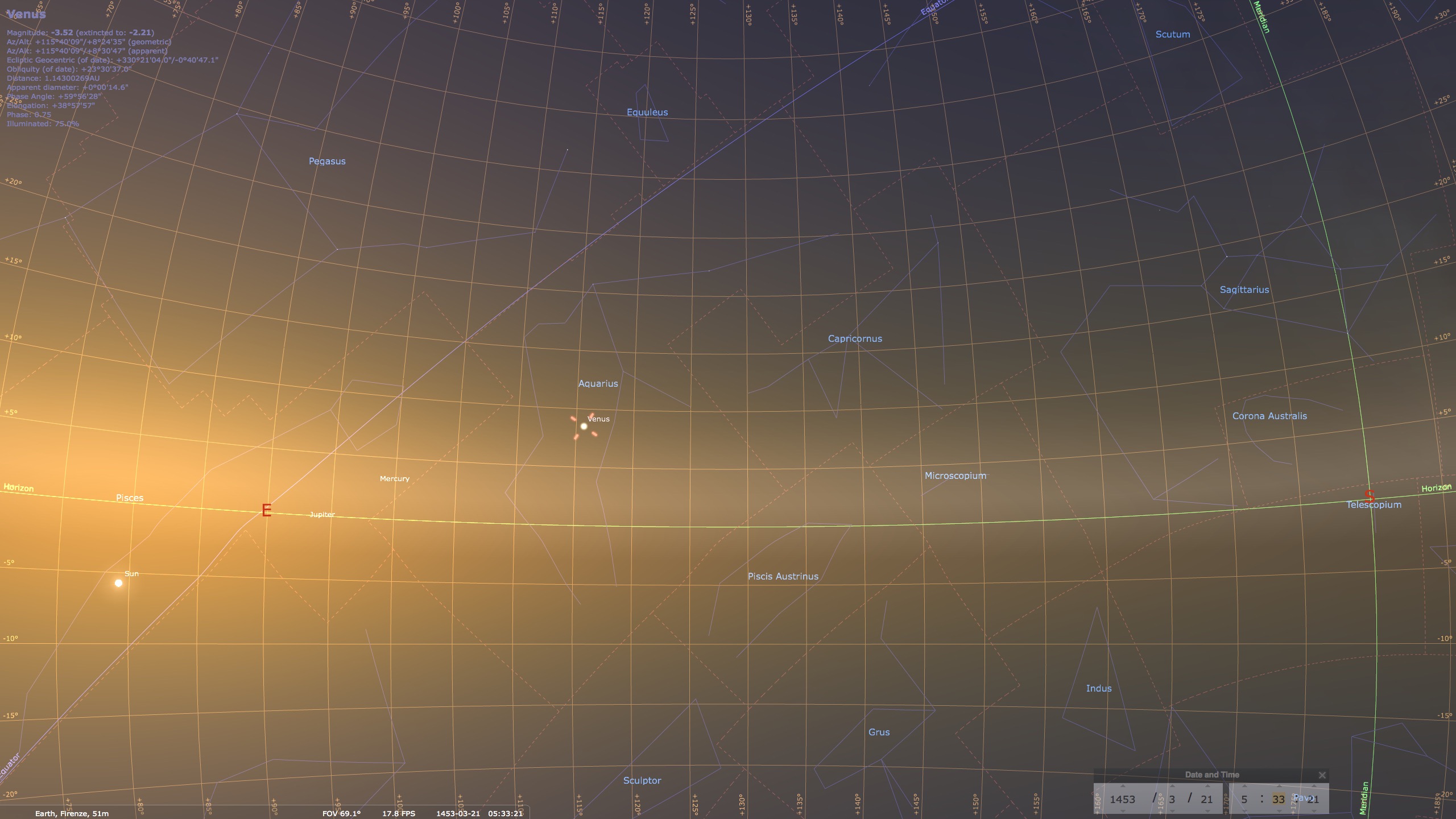}    
   \end{minipage}
\hfill   
\begin{minipage}{1.0\textwidth}
   \centering
   \includegraphics[width=\textwidth]{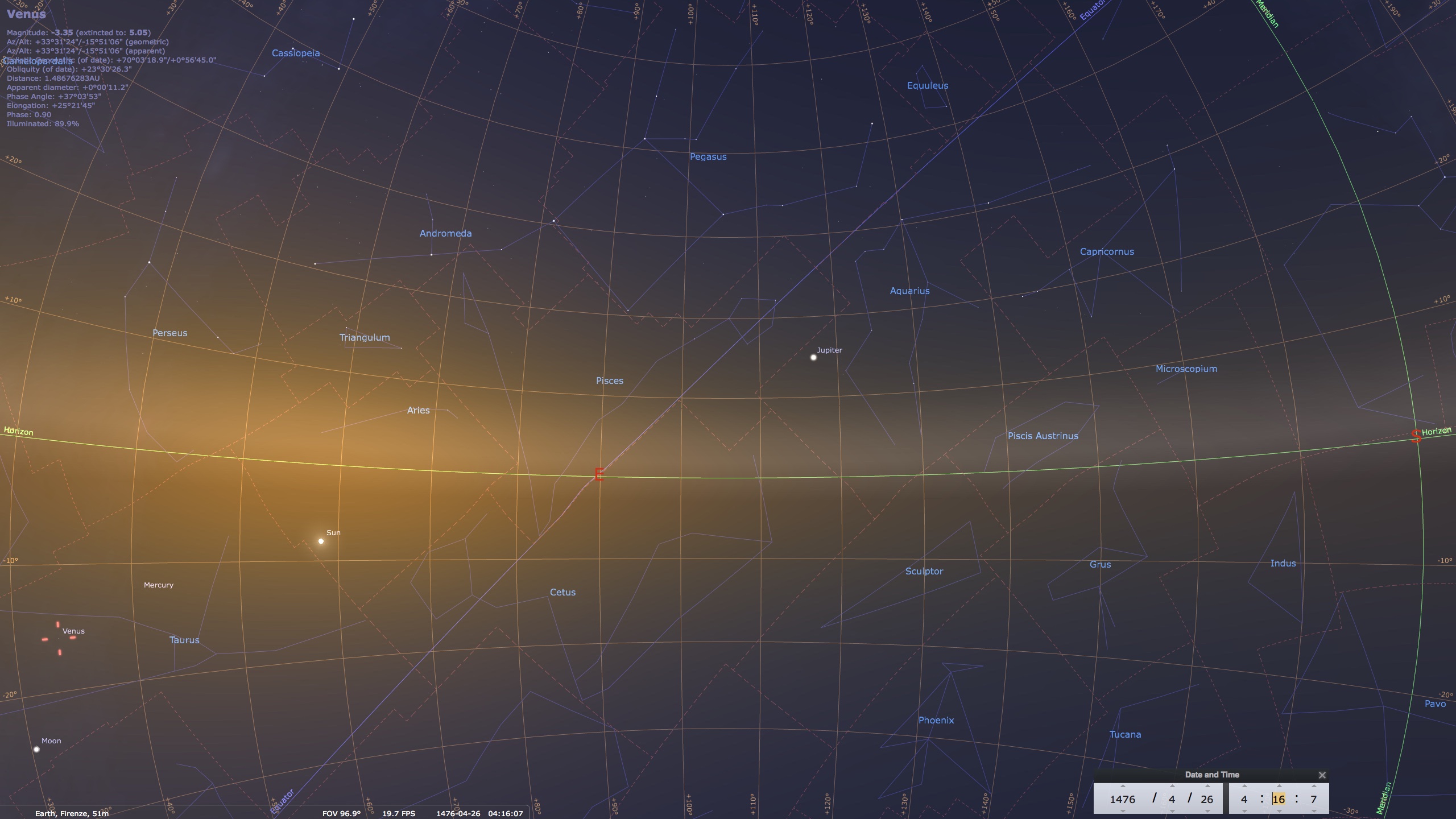}    
   \end{minipage}
 \caption{\label{Fig5}The sky on the spring equinox of 1453 (Venus is in conjunction with Aquarius) and on April 26, 1476 (Jupiter is in Aquarium and Venus appears at sunset).}
\end{figure}

The astronomical correspondences at the various dates mentioned in this paragraph prove the celestial and astrological themes that Botticelli might have pursued. Moreover, the reference to <<contending with Febo>> in Lorenzo's verses suggests an heliacal rising of the star if one hypothesizes that Simonetta is the planet Venus, ready to climb on its moving chariot.
As deduced by consulting the astronomical software, Venus is a constant theme in the dates related to Simonetta Vespucci: the planet Venus is still present at dawn of her supposed birthday and together with the planet Mars it moves along the ecliptic at the dates of her marriage, as well as from January (Lorenzo's Carousel month) until February 1469 (in conjunction with Mars in Sagittarius); towards the end of February Venus passes in Capricorn, then it swaps its position with respect to Mars and has a heliacal rising on the spring equinox in the Aquarius constellation next to  Capricorn that hosts Mars (figure \ref{Fig5}). 
This configuration lasts until the begging of April, while at the end of this month Venus disappears  just after the sunrise and Mars reaches Aquarius. Towards August 1469 Venus is almost always in conjunction with the Sun.
On April 26, 1476, the date of her death, we find Jupiter to the East of Aquarius close to the Sun and Venus at sunset. 

\paragraph{The Aquarius Stars as Satyrs, Botticelli's signature, Venus' right leg and jewel.} 
The most visible part of Aquarius - a not so bright constellation - is formed by a group of four Y-shaped stars representing the amphora from which the water comes out.
A good part of stars in Aquarius have names beginning with << Sad >> in Arabic meaning << fortune >>.
The main ones are the following: Sadalmelik ($\alpha$), the lucky stars of the king (from sa'd al-malik); Sadalsuud ($\beta$ ), the luckiest of lucky stars (from sa'd al-su'ud); Sadalbachia ($\gamma$ ), the lucky stars of the curtains or hidden things, hideouts (from sa'd alakhbiya); Skat ($\delta$ ), leg or tibia (from as-saq); Albali ($\epsilon$ ), devourer or the lucky one of the eater or the one swallowing; Ancha ($\theta$), lip (from latin); Situla (kappa),  bucket (from latin); Zeta Aquarii ($\zeta$), at the center of the letter Y, delimited by $\pi$, $\eta$ e $\gamma$ {\it Acquarii}.

As mentioned above before the Bayer Catalog the anatomical parts of the human figure were associated with the stars of the constellation and their disposition freely adapted for the interpretation that was intended to give. Traces of this association were found still two centuries later \cite{cirella}.  Therefore Skat and Ancha stars seem a hint for an explicit reference to the lower limbs:
<<So you have to remember that..[omissis]...Capricorn rules knees; Aquarius legs and shins>> ({\it De vita coelitus comparanda}, M.Ficino).
An unsolved enigma in {\it Venus and Mars} is the disappearance of the lower half of the right leg of Venus in the folds, perhaps accentuated to cover an anatomical error. Nevertheless, if we associate Venus to the constellation of Aquarius, then the representation of the constellation as reported in Igino's tables  - which Botticelli could consult or be aware of - where its right leg is visibly bent, also provides an explanation of its absence in painting NG915, rather than being a mistake made by the painter. Indeed a technical inspection conducted by the National Gallery failed to put such possible error in evidence \cite{davis}.
Moreover, almost all of the images of Aquarius from the Middle Age until 1482 are mostly dressed. Therefore, this could explain why the goddess does not show itself naked as the mythological encounter between Mars and Venus would require. Actually she represents a `celestial' Venus. 

While, on the one hand, we can assert with certainty that theta and delta stars can refer to the anatomical details of Venus' figure, on the other hand, if we consider the two constellations of Aquarius and Capricorn as a whole, the spear would disperse along the pouring water to the joining point with Capricorn. Then, Sadalmeik could represent the satyr with the helmet, Sadalsuud the satyr in the middle - in fact he looks at Venus << the most lucky of the lucky stars >>  - and Albali  << the devourer >> is  the last satire with the tongue out, especially if the fruit he holds in his hands corresponds to the {\it Ecballium elaterium}, much used as a purgative medicine.

And remarking that Lorenzo il Magnifico in {\it Simposio} wrote jokingly about Botticelli's greed: << Botticel whose fame is not blurry, Botticel I say; Botticello is more greedy than a fly, oh how many babbles of his I remember, when he is about to be invited to a have a dinner, one does not get in time to open the mouth that he already rushes dreaming of the food. Botticello goes and comes back full as a barrel>> \footnote{Author's translation from:<<Botticel la cui fama non \`e fosca, / Botticel dico; Botticello ingordo / Ch'e pi\`u impronto e pi\`u ghiotto ch'una mosca. / Oh di quante sue ciancie hor mi ricordo, / Se gli \`e invitato \`a desinar, \`o cena, / Quel che l'invita non lo dice a sordo. / Non s'apre a l'invitar la bocca a pena, / Ch' e' se ne viene, e al pappar non sogna, / Va Botticello, e torna botte piena>> }, we can affirm that the satyr  inserted in the armor is a mocking reminder of Lorenzo's verses, then a signature of Botticelli, located on the right side of the painting, as it is in the {\it Adorazione dei Magi}. Curiously, the only damaged part of the painting is just the little satyr's face inside the armor \cite{davis}.

Finally, Sadalbachia, <<the lucky stars of the hidden things>>, could be compared to the jewel worn by Venus: 8 pearls (symbolism of the bride and Venus?) surrounding a red stone (symbolism of love \cite{zoeller} or a reflection of the light from Mars?). 
The eight pearls could refer to Venus' visibility periods, although at Botticelli's time there were no known phases (it was Galileo to observe them in 1610 with the telescope as the effect of the planet's revolution around the Sun). 
In fact, for about 8 months the planet is visible to the West, then it disappears for seven days when it reaches its minimum distance from Earth (lower conjunction). Then it appears again to the East and remains visible for another 8 months. After this period Venus disappears for three months, totally illuminated by the Sun, on the opposite side, in the upper conjunction (maximum distance) with Earth. Then the cycle resumes. In addition, the path of Venus along its orbit draws against the Zodiac - as seen from Earth - a stellar pentagram (a pythagorean {\it pentalfa}?) every 8 years.

It is also worth remembering  that in Aquarius there are three meteor showers: {\it Eta Aquaridi} (maximum of 40 meteors per hour on May 5), {\it Delta Aquaridi} (twenty meteors per hour around July 28) and {\it Iota Aquaridi} (maximum of six meteors per hour on August 6). In the light of the interpretation presented here there is also the doubt that, in the absence of timely historical feedbacks, it was one of these transient objects to be seen by Lorenzo il Magnifico during his night walk and described in the sonnet for Simonetta.

\section*{Capricorn constellation in NG915 and the celestial gate}
Known in the ancient times and in Mesopotamia as goatfish (Suhur-Mash-Ha), Capricorn represented the god Ea destined to become later one of the three components of the Triad of creation, along with Anu and Enlil.
 
When the sign hosted the summer solstice, the Capricorn was called, on the Sumere cuneiform inscriptions, <<Father of Light>>  and was revered as a protector of men since he was thought he had saved them from the Universal Deluge by communicating to one of them, the Great Wise, the secrets for building an ark. Also called {\it Oannes} (greek version of {\it Eaganna}, namely  <<Ea the Fish>>), Ea was the god who had taught human beings the original doctrine.
Then the Greeks called him <<Egocero>> (horned goat) and identified him with Pan, the god of the countryside with horns and goat's legs. And this would explain the presence of the Satyrs with horns and paws in the encounter between Venus and Mars instead of the {\it Amorini} predicted in the mythological story.  
In support of such an assertion, the Mars planet at the vernal equinox of 1482 was in Sagittarius. Eratosthenes described Sagittarius as a two-legged creature with a tail of a satyr called {\it Crotus}, rather than the traditional Centaur.
  
According to Igino, Crotus's father was Pan and, according to the myth, Pan was a playful creature from uncertain birth, intended for most of the time to hunt women or to sleep. His gift was to scare people by the particularly strong shout from which the word <<panic>> originated.
Pan rescued gods twice: the first one during their fight against the Titans by blowing in a shell and scaring of the enemies (according to Eratosthenes the connection with the shell explains the partial fish representation attributed to the Capricorn); the second one, when by screaming he warns the gods of the approach of Typhoon. 
Thank to his services, Zeus raises Pan in the sky as the constellation of Capricorn.
 
It is well known the astronomical meaning of this story: Zeus represents the Sun that at the winter solstice remains at the mercy of the darkness from which it can only escape thanks to the `goat-fish' solstice, point in the sky of great symbolic value at which the inversion of the path of the Sun towards to spring takes place. A kind of heavenly gate and a symbol that reappears, for example, in the Gospel according to St. John and in the {\it Sol Invictus} celebration 
\footnote{At the time of Eratosthenes and Hipparchus, the winter solstice was still in this constellation, but because of the precession of equinoxes, today the winter solstice falls into Sagittarius.}.
\begin{figure}
\begin{minipage}{0.3\textwidth}
  \centering
\includegraphics[width=\textwidth]{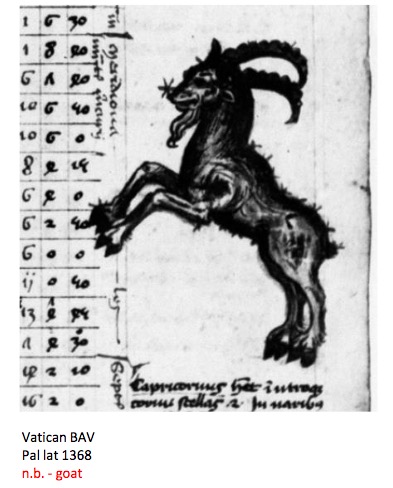}
\end{minipage}
\begin{minipage}{0.3\textwidth}
 \centering
 \includegraphics[width=\textwidth]{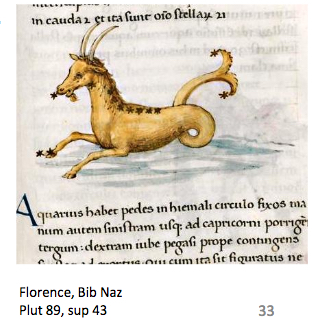}
   \end{minipage}
\begin{minipage}{0.3\textwidth}
\centering
\includegraphics[width=\textwidth]{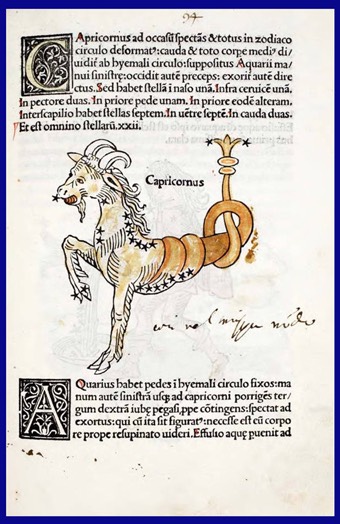}
\end{minipage}
 \caption{\label{Fig9} On the left Capricorn as a goat in `Ptolemaic' stellar tables, Vatican BAV Pal lat 1368 \cite{lippncott}; in the middle  Capricorn in {\it Germanicus  Aratea,  Hyginus,  Astronomica}, Biblioteca Laurenziana, Plut. 89. sup 43 \cite{lippncott}; 
 on right depiction of Aquarius in the publication of {\it Poeticon}, 1482.}
\end{figure}
\hfill

In figure \ref{Fig9} one can analyze the images of Capricorn that Botticelli might have seen. Note the position of the right knee in the {\it Poeticon} edition of 1482, very similar to that of Mars in NG915.

Bode \cite{bode} and Schmarsov \cite{schas} already suggested that NG915 refers to the Medici Carousels.
Lorenzo il Magnifico was born on January 1, 1449, just under the sign of Capricorn. The Medici family assumed this constellation as a regal and power identifier. Even the helmet used on February 1469 during the Carousel (<<elmetto fornito d'ariento con un Marte per cimiero>>) referred to the birthday theme of that day: on the front was painted Mars (considered the dominant planet of Ascendant Scorpio) resting his feet on an eagle  (<<L'aquila rossa in sull'elmetto un Marte sopra sua stella fe' d'argento e d'oro>> ) while on the back there was a goat's head, probably depicting the Capricorn Zodiacal Sign. 
This would explain why the first satyr that holds the spear - Aquarius's alpha or << the lucky stars of the king >> - wears a helmet whose front looks to Mars and the back, Capricorn, to Venus (like the alignment in the sky). A clear reference to the rise of Lorenzo, who remained miraculously alive on the day of the Pazzi Conspiracy, while his brother Giuliano died (the Sunday before ascension). 

\paragraph{Giuliano dei Medici as the planet Mars.}
Angiolo Poliziano depicted Giuliano dei Medici as the following <<He was of high stature with a well-proportioned body, with large and protruding pectorals. He had turned and muscular arms, flat belly, lively eyes, indomitable face, and long black hair>> \footnote{Author's translation from <<Fu di alta statura un corpo ben proporzionato con pettorali ampi e sporgenti. Aveva le braccia tornite e muscolose, il ventre piatto, occhi vivaci, il volto indomito, capelli lunghi e neri>>.}.
According to the chronicles Giuliano dei Medici fell in love with Simonetta Vespucci. He was born like Simonetta in the year 1453, on October 25th, with an heliacal rising of Mars, not so much visible, and in conjunction with Saturn. The association with the planet Mars makes more plausible the red hair of the figure representing him in NG915.  

The exaltation of their history took place in the Carousel fought on January 29, 1475 in Piazza Santa Croce to celebrate an important peace agreement between Florence, Milan and Venice realized by il Magnifico. Giuliano dedicated the victory to Simonetta who was present among the people crowding Piazza Santa Croce. A banner - conceived by Poliziano - was commissioned to Sandro Botticelli for Giuliano, where Simonetta was depicted in the allegorical dress of Venus-Minerva with a Cupid tied to her feet. The motto <<La sans par>> on the banner was personally chosen by Lorenzo. 
Like Luigi Pulci who dedicated a poem years before to the Lorenzo's Carousel, Poliziano wrote the {\it Stanze} for Giuliano, emphasizing Giuliano's fight was devoted to his love, the beautiful Simonetta Cattaneo Vespucci. It was interrupted because of the Giuliano's death.

Giuliano, in fact, was killed in the Pazzi Conspiracy two years later, on April 26, 1478, indeed dying on Simonetta's day of death, while Lorenzo was injured but not severely. Giuliano was buried in a uncovered coffin like Simonetta. Coincidences that perhaps could not go unnoticed in the neoplatonic environment. In this regard, critics Enrica Tiezte-Conrad (1925) \cite{conrat} and Carlo Gamba (1936) \cite{gamba} already identified in the layering and contiguous figures of NG915 a similarity with Etruscan funerary representations\footnote{Tiezte-Conrad referred to a detail of the cover of a third-century sarcophagus (in the Vatican Museums) depicting two female figures reclined in contraposition against a Dionysian scene.}.
\begin{figure}
   \centering
\begin{minipage}{1.0\textwidth}
   \centering
   \includegraphics[width=\textwidth]{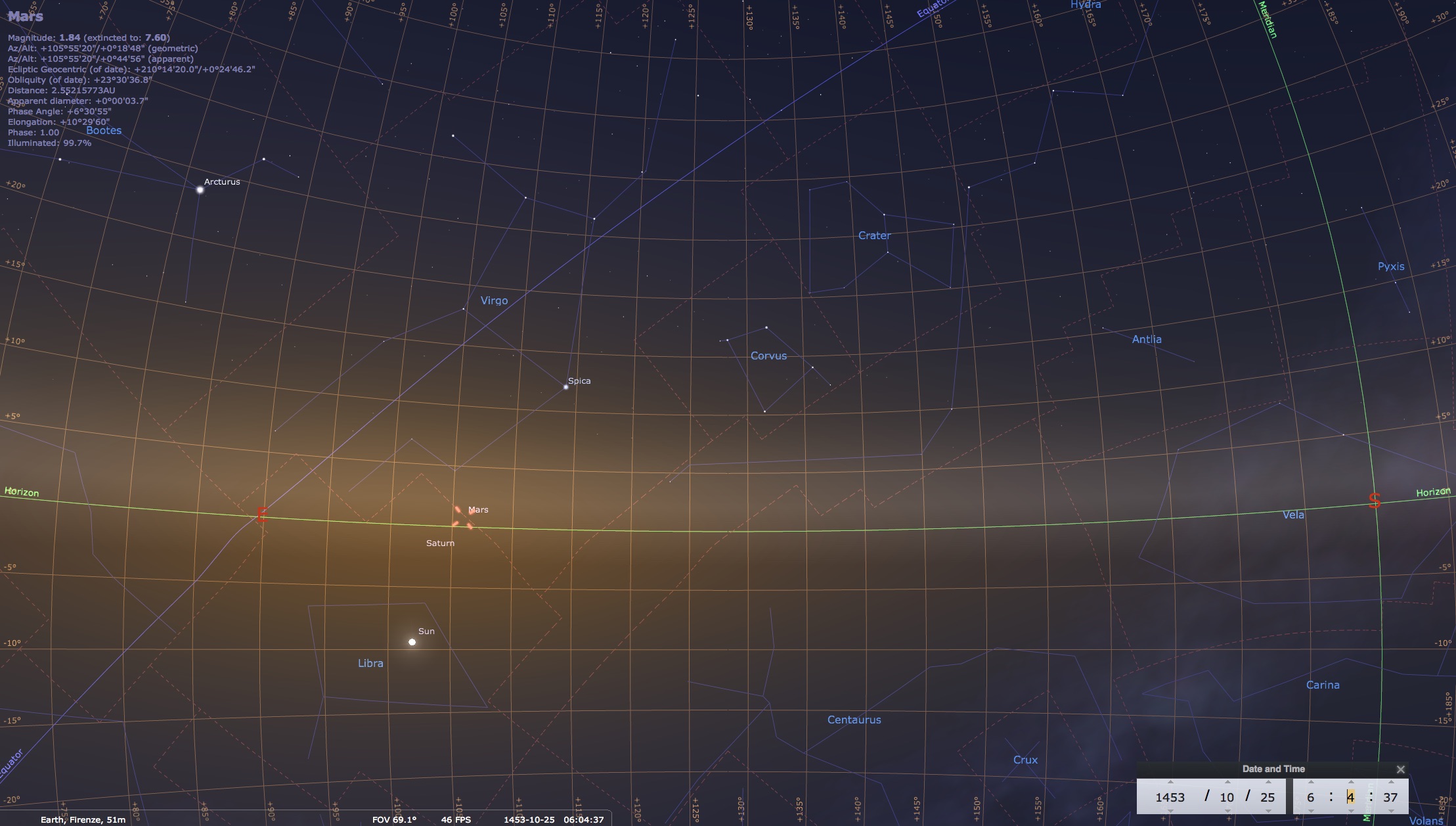}    
   \end{minipage}
\hfill   
\begin{minipage}{1.0\textwidth}
   \centering
   \includegraphics[width=\textwidth]{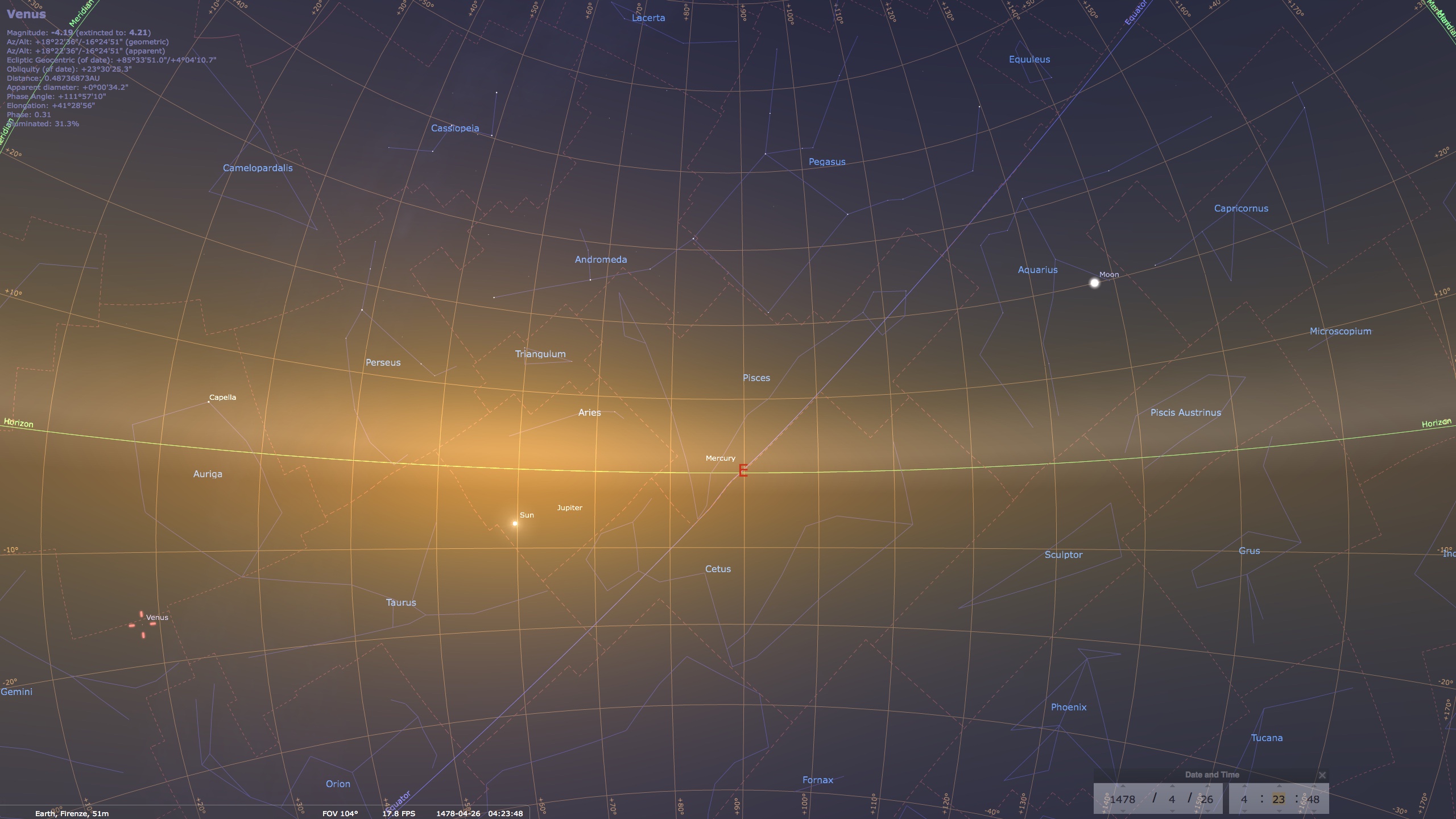}    
   \end{minipage}
\caption{\label{Fig8}Astral Situation, in Florence, on 25th October 1453 (heliacal rising of Mars) and on 26th of April 1478 (Jupiter is in conjunction with Sun and Venus appears at sunset).}
\end{figure}

\paragraph{The NG915 marine shell and the sacrifice stars of Capricorn.}
 
The brightest stars of Capricorn are located at the border with Aquarius to form a sufficiently recognizable triangle, whose vertices are {\it Algedi} to the North-West, {\it Deneb Algedi} to the North-East and $\psi$ Capricorns to the South.
Let us list them: Algedi ($\alpha$) o Giedi,  the kidskin (from arabic al-jady), or two stars, alpha1 e alpha2, called {\it Prima Giedi} and {\it Secunda Giedi}; Dabih ($\beta$),  the luck of sacrifice or the butcher  (from arabic al-sad al-dhabih); Nashira ($\gamma$), the lucky one or <<the bearer of good news>> (from arabic); Deneb Algedi ($\delta$), to the tail of the kidskin (from arabic al-dhanab al-jady);
Alshat ($\nu$), the slaughterer's lamb,  meaning the sheep that was to be slaughtered by the adjacent Dabih.
The names are related to the Id-al-Adha feast celebrated among Muslims in the period of the winter solstice, exactly on the day when the moon was between Dhabih and Alpha-Beta of Capricorn, during which goats were sacrificed to promote healing from diseases.
It would appear, therefore, that the third satyr that blows in the marine shell - an important part of the myth of Capricorn and also present in that of Aquarius with reference to Triton  - is associated to Nashira, {\it gamma Capricorni} located at the border with Aquarius.

The link in NG915 could be at the failed Pazzi Conspiracy, considering that thanks to the "astral" sound of the seashell, Firenze was saved from the danger of enemies' occupation.  The association with the stars of Capricorn is even more relevant if one thinks of the meaning of their names related to the concept of sacrifice and that Giuliano, as matter of fact, was the only Medici to be ``sacrificed'', the first to die hit by many stabbers.

Poliziano tells that Lorenzo il Magnifico was first grabbed by his shoulder and then hit at the throat, but he could shield from the following strokes with his arm to which he had wrapped his cloak and then, once freed, defended himself with his sword. Lorenzo's fortune was due to the fact that his attackers were two inexperienced priests, who were chosen as slayers only at the last moment. In fact, initially the chosen hit man was the leader Giovanni Battista of Montesecco who, after a rethinking, refused to commit homicide in a sacred place (the Conspiracy took place in the church of Santa Maria del Fiore). 
Indeed Lorenzo was surrounded by his friends and escorts, and his friend Francesco Nori died in the attempt to defend him from the killers. 
On the body of Giuliano acts cruelly Francesco dei Pazzi (and also Baldini)\footnote{<< Con tanto studio lo percosse, che obcecato da quel furor che lo portava, se medesimo in una gamba gravemente offese>> (N. Machiavelli)}. 

Like Pan, Giuliano (or Iulius in the Poliziano's {\it Stanze}) ascends to the heavens. The hilt indicated by the left middle finger of Mars' hand confirms that `` I '' stands for ``Iulius'', as has already been recognized by Bellingham \cite{Bellingham} and Guidoni \cite{guidoni}.

\paragraph{Lighting from Mars, light in the sky, and the plants in the astral winter of NG915.} 
Assuming that the picture represents the rising Sun, the light of dawn spreads on the background of NG915. Mars-Giuliano is therefore directly illuminated by the Sun (unlike Venus turning her shoulders to it, the elongation of the planet is twenty degrees). Perhaps the satyr with the shell is working to wake him up, warning that a new astral life is coming.
Associated with the nature of Winter, Capricorn as a sign of earth symbolizes the seed that begins its slow and progressive maturation until it reaches rebirth on the surface in Spring. Bellingham was the first to suggest that the absence of flowers is due to the winter setting of the Carousel \cite{Bellingham}. In a recent publication Paoli reports that the vegetative stage of plants in the painting is typical of the months preceding Spring \cite{paoli}.
In addition to the {\it Ecballium elaterium} in NG915 we find: on the  bottom right the {\it Plantago lanceolata}, called also Mars grass (<<{\it Herba quarta Martis armoglossa dicitur}>>, {\it Albertus  Magnus}), used as a remedy for women and by Greeks called lamb's tongue \cite{mirella}; close to the hilt the {\it Poterium sanguisorba L.}, used in ancient time against hemorrhoids and in salads; finally close to Venus {\it Tragopon} that in Greek means goat beard\footnote{For more details the reader can refer to the book by Mirella Levi D'Antona\cite{mirella}}. This emphasizes their connection with the stars of Capricorn and the interpretation of the fourth satyr as Albali in Aquarius. 
We can also assume that in the background there is the myrtle and the {\it Laur Nobilis} (not flourished), the first one sacred to Venus and the second one symbol of conjugal love \cite{mirella}, but at the same time linked to Lorenzo il Magnifico and Pierfrancesco di Lorenzo (see below) because of the assonance with their name and the ancient symbology of power: <<E 'l laur che tanto fa bramar sue fronde>>,  <<E tu ben nato Laur; sotto el cui velo Fiorenza lieta in pace si riposa>>({\it Stanze}, 82.4 and 4.1) \footnote{Note that the spring equinox in the calendar of that epoch fell around March 10, the absence of blossoms may be justified as NG915 symbolically represents an astral ascending.}.
 
\section*{Marsilio Ficino and the stellar harmony}
The role of Marsilio Ficino in Florence is known and widely studied. However it is worth highlighting some aspects of his philosophy in support of the astronomical interpretation given to NG915. 

Marsilio Ficino had spread the ancient mythologies and the works of ancient poets - in particular Homer, Orazio and Ovid - and together with Poliziano he was master of Lorenzo il Magnifico who was later patron of Botticelli. In 1459 Marsilio Ficino created the Platonic Academy which sought to reconcile Platonic ideas with Christianity.
According to many critics \cite{gombrich, kps, boskovitz, ferruolo, robb}, NG915 should be read in harmony with the philosophical themes expressed by the Neoplatonic Academy, in particular the ones of marriage. The painting is the representation of the encounter between Venus (depicting ``catastematici'', i.e., static, pleasures) and Mars (representing dynamic pleasures), a concept found in the proemio of {\it De rerum natura} by Lucrezio.
In addition, Olson \cite{olson} assimilates the figures of Eve and Adam in the upper frame of the gate of Ghiberti's paradise with those of Venus and Mars in Botticelli's painting, both as regards the contrast between the dressed woman and the naked man, as well as for the postures that recall the ancient fluvial divinities, as proved in the present work.

In the {\it Symposium} Marsilio Ficino writes about the harmony of opposites symbolized by the duality of Mars-Venus and the superiority of the goddess Venus on the god Mars\footnote{ <<If you fear Mars put against Venus, namely Venus is in harmony with Mars>>  M.Ficino, author's translation from <<Se temete Marte opponetegli Venere, ovvero Venere \`e in aspetto armonico a Marte>>.}. Venus symbolizes {\it Humanitas}, love and concord, while Mars hatred and discord, so much so that the ancients recognized in him the god of war.
Living in the Beauty helps to overcome the earthly dimension, and Venus, linked to the concept of beauty and contemplation, is a symbol of spiritual elevation through the arts, nature, knowledge and love. The Creation is only possible through love: Love is therefore at the foundation of the cosmos, and only through love the laws of the universe are understood and one is able to approach God.

Inspired by Plotin \cite{canaglia,faggin}, Ficino states that pure philosophy (the one devoted to God's devotion) has to deal only with tho se sciences that allow admiring the predetermined paths of the stars subjected to numerical laws.
But the universal harmony (<<universal sympathy>>) for its very essence reverberates on all the levels of being by investing them with its own law and order, as a magical plot that envelops the whole reality by establishing a strong causal network of  invisible constraints. Harmony is born from the << opposites >>, but also by the << similar >>, as all things are related. The knowledge of the qualities and virtues of planets, constellations, and zodiac signs, but also of the time and progression of the celestial configurations, helps to recognize the bond of concordance that connects to them the acts and the inclinations of men \cite{faracovi-1,faracovi-2}.

In addition, NG915 can also be read as a symbol of Peace that demolishes the War of Weapons, to be linked to the new Florentine climate created by the diplomatic abilities of Lorenzo il Magnifico after Pazzi's conspiracy, as already assumed by Cheney \cite{cheney}.
It is worth noting that Ficino studied also the works of Ya'qub Ibn Ishaq al-Kindi, philosopher, scientist and theoretician of the magic arts, who lived in the 9th century and wrote the {\it De Radiis}, one of the most widespread magic manuals in Western europe \cite{corbin}. Chapter IX of the {\it De Radiis} is devoted to animal sacrifices: a dying animal naturally fits in the ordinary concatenation of the events; the intentionally killed animal, on the other hand, temporarily disrupts a balance, it comes to form an opposition to the course of nature which, if altered, opens a gap in the order of the real and creates interference zones that stop the ordinary flow of things. Al-Kindi speaks of the sacrificial act intended as the creation of a << natal theme >> \cite{melis,katins}. 

But as there is a macrocosmic order between the forces that govern the world and whose cycles are qualitatively similar to human ones, there is likewise a microcosmic astronomy:  imagination is an << astro >>, which creates a causal  << astral >> impulse, like a seminal power that can only be activated in virtue of the faith and the intentional strength that the operator, the {\it alter deus}, is able to put in action, i.e. through the psychic life, acting at the same time on the quality and  <<signatur\ae>> of nature.


Marsilio Ficino not only translated {\it Hermes Trismegisto}, but also elaborates his own magical practices in accordance with Christian neoplatonism. 
As Hermes Trismegisto Ficino puts man into the physical center of the world, conceived as a unity and totality. The symbolism of magic arts was vast, and it is found on the ancient astronomical charts that contained, besides the zodiac, a myriad of strange figures each of them associated with a planet, star, or constellation \cite{garin,treccanimagia}. 
The quality of a ``magician'' was considered in direct relation with his ability to grasp the symbols and relationships between the things of Earth and those of Heavens \footnote{ <<Nature it is not an inanimate house, but it's all covered by attractions and repulsions. The magician is the one who, knowing the sympathies and the contrasts, and generally the quality of constraints, is able to act on them and to connect similar things as the farmer marries the elms to the screws>>, author's translation from <<la Natura non \`e una casa inanimata, ma \`e tutta percorsa da attrazioni e repulsioni. Il mago \`e colui che, conoscendo le simpatie e i contrasti, e in genere la qualit\`a dei vincoli, \`e in grado di agire su di loro e di collegare cose simili, come l'agricoltore sposa gli olmi alle viti.>>({\it Disputatio contra iudicium astrologorum}, 1477).}.

\paragraph{The marriage in NG915.}
The size and shape of the painting suggest that the work has been commissioned for a wedding, more exactly to be placed on the cover of a hope chest.
On July 19, 1482 Lorenzo di PierFrancesco dei Medici (called Popolano) and Semiramide Appiani (nephew of Simonetta) married \footnote{Simonetta's marriage is also assumed to have been celebrated towards August.}.
This union was asked by Lorenzo il Magnifico, who had previously thought of marrying his young brother Giuliano with Semiramide because he aimed at the iron mines on the island of Elba in possession of the Appiani family (note that iron is also the symbol of Mars). The marriage of a girl Appiani with a Medici man revived the myth of Simonetta, to whom Semiramide apparently resembled.

Lorenzo di Pierfrancesco dei Medici was born in Florence on August 4, 1463 and at the death of his father on July 19, 1476, was placed under the patronage of Lorenzo il Magnifico and Giuliano. His teachers were Giorgio Antonio Vespucci, Marsilio Ficino, Naldo Naldi, and Angelo Poliziano \cite{biotreccani}. Taking advantage of his role, in 1478 Lorenzo il Magnifico took possession of more than 53,000 florins cash from Pierfrancesco's inheritance to cope with the crisis that had hit the Roman branch of the Medici counter following the Pazzi Conspiracy.
Moreover, Pierfrancesco was acquainted with Vespucci family, especially with Amerigo Vespucci who dedicated to him his treatise {\it Mundus novus} (1504)\footnote{Amerigo Vespucci was sent  as agent to the branch in Seville at the service of Lorenzo di Pierfrancesco; during his stay in Seville he developed the idea to sail to the New World.}.
It was with Botticelli that Pierfrancesco  had a lasting relationship by commissioning to him various works. It seems that Pierfrancesco ordered {\it Primavera} and {\it Nascita di Venere} to Botticelli to adorn his bedroom in the Palace of Via Larga. The historians, however, do not all agree on the commission, and there are those who claim that the paintings were given to Lorenzo il Popolano by Lorenzo il Magnifico, who actually ordered the works, after the `Lodo Scala'. On the other hand, Lorenzo il Magnifico was also the client of the {\it Banchetto di Nastagio} at that time, painted by Botticelli around 1483.

The reference to Vespucci is also confirmed by the motif of the wasps in the upper right, highlighted for the first time by Gombrich \cite{gombrich}, quite unusual given the winter setting of the painting. Here, we emphasize that such insects are particularly active between July and August, presumed period of marriages we refer to and month of birth of Lorenzo Pierfrancesco.
On the other hand the quoted sonnets by Lorenzo il Magnifico represent a exegetical key for NG915, being a clear allusion to the real customer of the work. Along with the transfiguration of the tragic, political, historical and personal events according to the Ficinian 'astronomy/philosophy', in the opinion of the author, facetious aspects have to be considered as an invitation to enjoy the happy moments of life - such as a marriage - because:  <<Chi vuol essere lieto, sia: di doman non c'\`e certezza>>  (Lorenzo il Magnifico, "The Triumph of Bacco and Arianna'').

The iconography could therefore have been chosen as a wish by Lorenzo il Magnifico - who is widely represented in the compositional scheme - towards the bride and groom who, given the correspondence and affinity with Giuliano de Medici and Simonetta Vespucci, might be Lorenzo di Pier Francesco dei Medici and Semiramide Appiani.
The commitment of NG915 by Medici was also suggested by Salvini \cite{salvini}.

\section*{Astronomical <<signaturae>> as disambiguating keys for \textit{Venus and Mars} and Venus' mythological representation}

As is often the case with Botticelli's works, we are confronted with a symbolic truth expressed at many levels, with a thick plot between allegories and philosophical concepts, with a game between the ambivalence of the meanings and the real reality.

In a recent book \cite{paoli}, Paoli speculates that in NG 915 is painted the parody of a myth, the recipients of which were Simonetta and Giuliano: one would allude to the missing copula between Venus and Mars, to Ganymede, to the supposed masturbation of Mars typical of God Pan, to the scorn of the satyrs, to the resentment of the Vespucci family towards the Medici, to a negative opinion on the adulterous Simonetta by Savanarola's followers, and so on.

The present study is instead aimed at providing an astronomical reading of the pictorial composition while leaving a window open to the possibility of coexisting different levels of communication and interpretation.
A heavenly connection, though only of mythological and non-astronomical nature, had been mentioned by Langton Douglas \cite{douglas}, but later never fully and seriously investigated.

This preliminary analysis already highlights remarkable and non-negligible coincidences, such as, for example, the motion of the planets (reflected even on Venus' jewel) and the meaning of the star names with the characters and the stories that the NG915 painting invites us to consider. In particular $\epsilon$ {\it Aquarii}, Albali, can be read as a signature of Botticelli according to the verses wrote by il Magnifico about artist's greed. 

The heliacal rising, in particular, is clearly the thread by which it is possible to decipher Ficino's message and the historical events; in addition the conjunction of Venus and Mars with Aquarius and Capricorn provides the name and date, i.e. 1482, the same year of the first print publication of {\it Poeticon Astronomicon} by Ratdolt (in which Aquarius and Capricorn show similarities with the figures of Venus and Mars in NG915).  

Furthermore, specific astronomical phenomena such as the helical raising in conjunction of stars/planets could be the key to interpret some mythological figures, as the case of Venus which appeared quite often in heliacal rising in the so-called 
 `` Celestial Waters '' \\ region of the sky, i.e. where Acquarius is located,  during the spring equinox at the Mesopotamian latitudes. Indeed Venus belongs to the old pantheon: according to Esiodo was the daughter of Heavens and Sea, or Uranus and Gaia, and Ihstar was often depicted in conjunction with the Sun in the {\it Lucifero} aspect \footnote{A quick check with the astronomical software confirms this fact therefore deserving a specific attention in a next study.}. A clear evidence of the association Venus/Acquarius is, for example, the megalographic wall fresco from the perystiulium of the {\it Casa di Venere in Conchiglia} at Pompeii. 
 \begin{figure}
\begin{center}
\includegraphics[width=12.0cm]{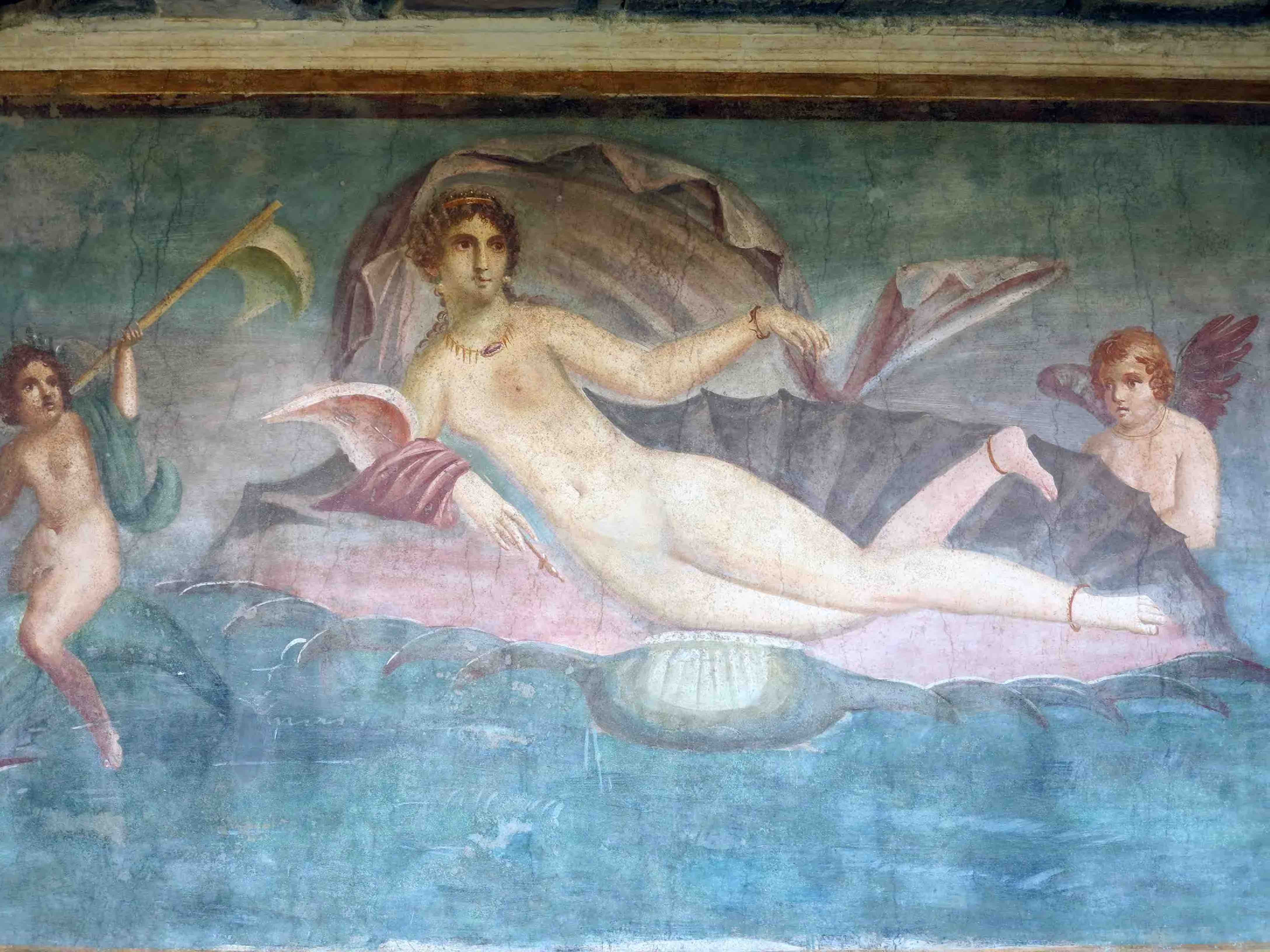}
\caption{\label{venere_pompei} Megalographic wall fresco from the perystiulium of the {\it Casa di Venere in Conchiglia} in Pompei.}
\label{default}
\end{center}
\end{figure}
The fresco shows one of the most scenographic example of Venus lying on a shell widely diffused in classical antiquity. Venus holds a fan in her right hand while the left holds the veil that swells with the wind as the water flow of Aquarius. The hairstyle shows the typical flavian curls. The goddess wears a diadem, a necklace, gold bracelets on the wrists and ankles. The garden was originally embellished with myrtle plants and gallic roses. The legs are crossed according to a typical pattern, the right leg and the fingers of her left foot show at first glance evident similarities with those of Venus in NG915. The residential {\it  domus}, located in the amphitheater quarter, was brought to light only in 1952. However the strong coincidence with the Venus in NG915 is impressive, suggesting that Botticelli during his stay in Rome just before 1482 saw this iconography of Venus in some roman collections.
  
\section*{Conclusions}
The analysis exposed in the present essay can be considered  the basis for further insights by specialists from the various multidisciplinary sectors (historical, philosophical, artistic, exegetical, literary) necessarily brought up by this astronomical study, hoping to open new targeted readings. 
\begin{figure}
\centering
\begin{minipage}{1.0\textwidth}
\includegraphics[width=\textwidth]{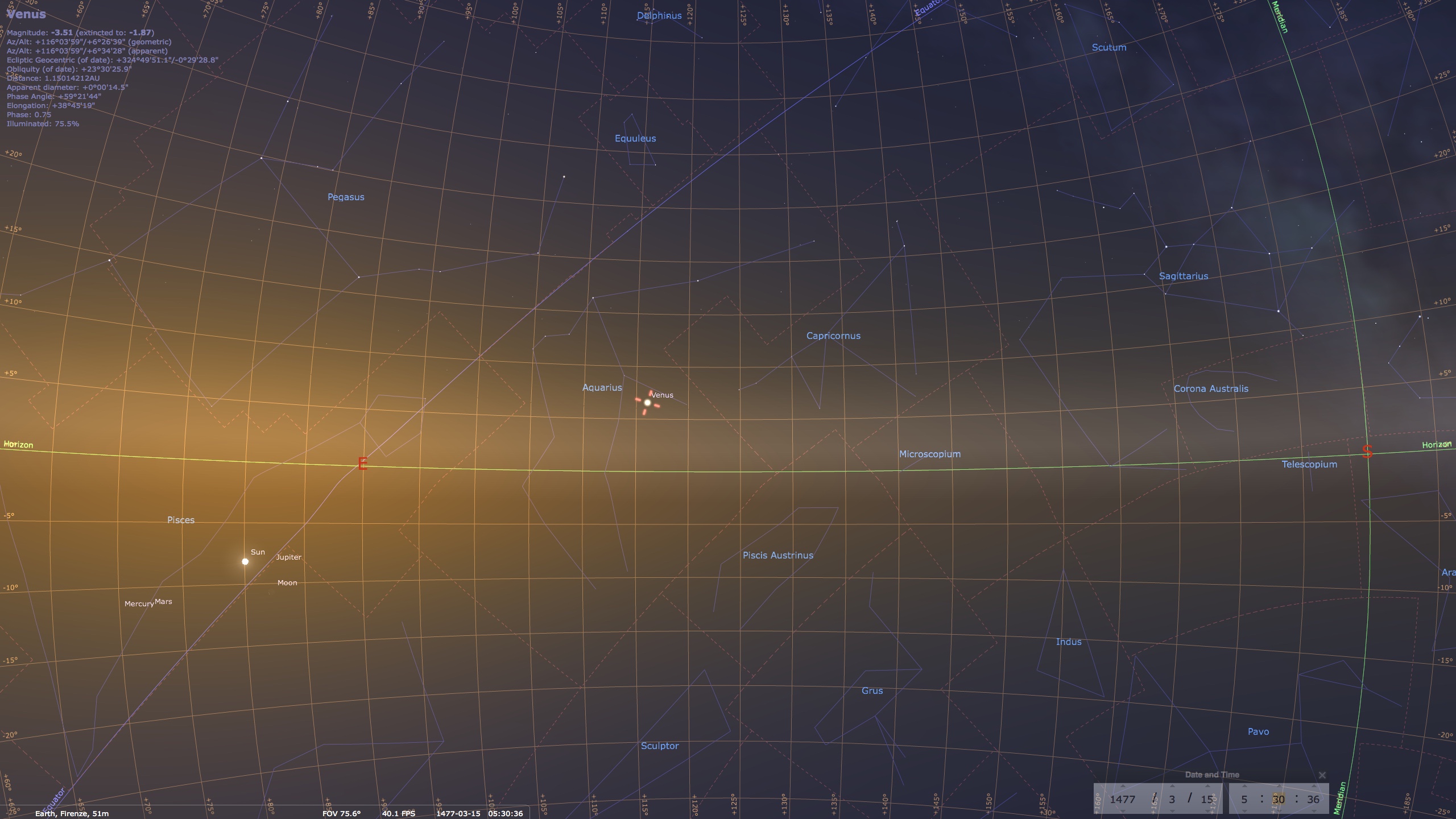}
\end{minipage}
\hfill
\begin{minipage}{1.0\textwidth}
\centering
\includegraphics[width=\textwidth]{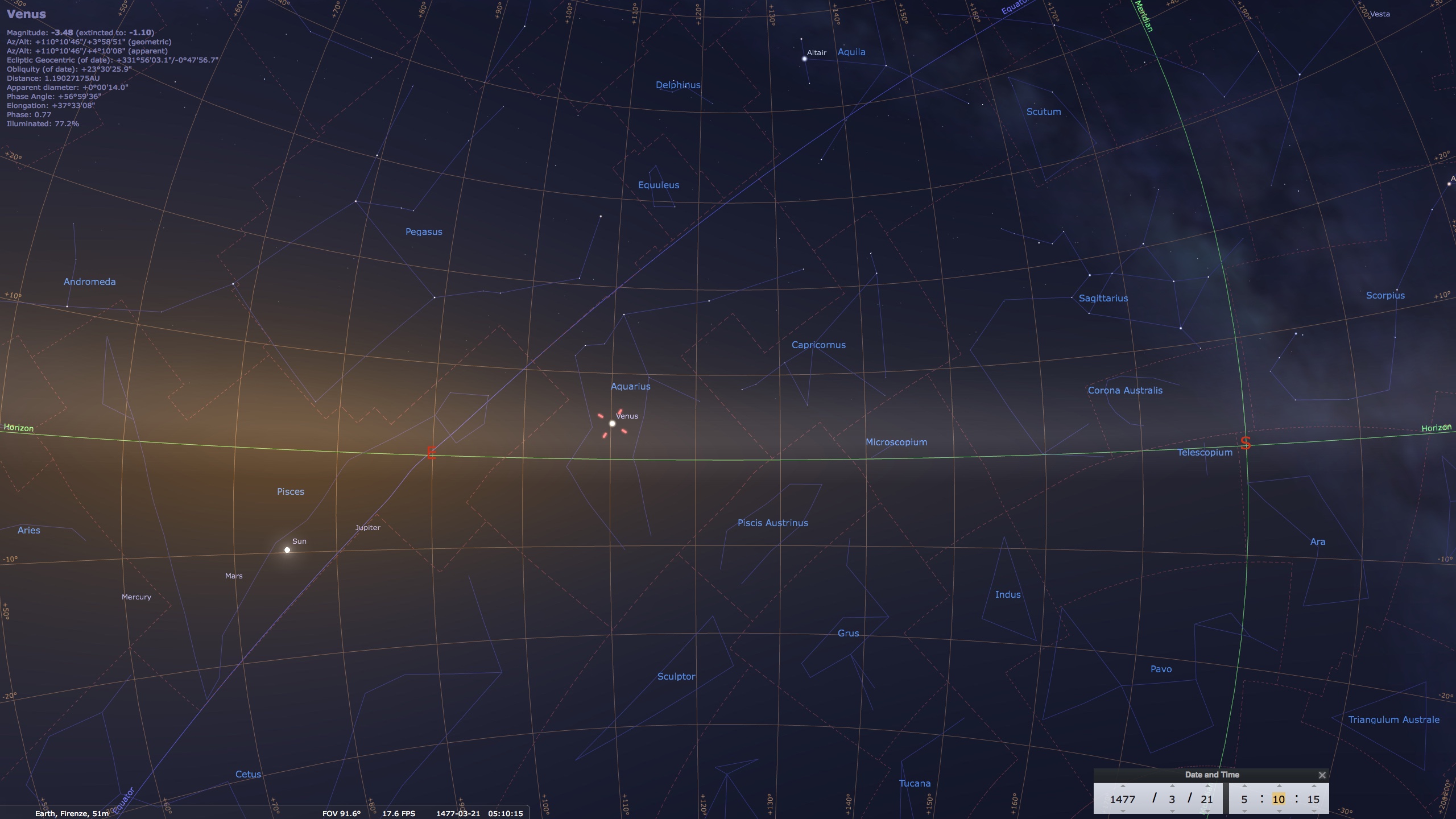}
\end{minipage}
\caption{\label{Fig10}Stellar configuration, at Florence, around the vernal equinox of 1477.}
\end{figure}
For example, it would be desirable to have a more in-depth study on the astrological (classical) aspects concerning the real positions of the planets, as well as on the celestial correspondences with the plants as set forth in Marsilio Ficino's treatise to explore further astronomical configurations and allegories \footnote{In the Enneade IV book (chapter 32) Ficino outlines the image of a Universe as of <<a living unitary, embracing the living all who are in it and is endowed with a unitary soul spread on all its parts>>, author's translation from <<un vivente unitario, che abbraccia i viventi tutti che son nel suo interno ed \`e dotato di un'anima unitaria diffusa su tutte le sue parti>>.}. 

As, if we admit the astronomical interpretation, we confirm the intepretation that {\it Pallade ed il Centauro}, {\it Primavera} \footnote{Both were exposed above the entrance door of the antechamber in the Palace in Via Larga where Lorenzo il Popolano lived. In the {\it Primavera} painting Semiramide Appiani would have been recognized in the central figure of the Three Graces, to represent spiritual love, i.e. the {\it Humanus} Love in the neoplatonic philosophy; Pierfrancesco, on the other hand, would have been portrayed in Mercurio's clothes (and those of Mars, since he wears also a sword).}, and {\it Venus and Mars} may have been conceived of representing first the absolute victory of Love on brutality and lower instincts (the Centaur), thus as a regenerating force thank to fertility that in April revives the Earth (Spring) and, finally, as contemplation of Love, namely the ascension of the soul to Heavens in the harmony of opposites (Venus and Mars). 
Better yet, if we include also {\it Nascita di Venere}, so the works as a whole celebrate the cyclical rhythm of being through the vital principle: birth, life and death.

Moreover, according to Ficino in {\it De Vita Coelitus comparanda}, Venus, Jupiter and the Sun are the three "stars" propitious to man associated with the Three Graces of Spring \footnote{ << The Three Graces are Jupiter, Sun and Venus. Jupiter is the Grace, which is between the other two and is particularly appropriate to us, >>  fifth chapter of {\it De Vita Coelitus comparanda}, while in the sixth chapter it is told that << Where it deals with our virtues, the natural, the vital and the animal ones, and which planets help and how they act it through the appearance of the Moon to the Sun, to Venus, and especially to Jupiter >>. Author's translation from <<Ove si tratta delle nostre virt\`u, quella naturale, vitale e animale, e quali pianeti le aiutano e in che modo lo fanno tramite l'aspetto della Luna al Sole a Venere e specialmente a Giove>>.}.
These three stars always come as gifts of Joy, Splendor and Freshness, which are precisely the translation from the greek names given to the Three Graces: respectively Euphrosine, Aglaia, and Talia.
In the wake of the analysis presented here, we report the image of the astronomical ephemerides of the spring equinox of 1477 in figure \ref{Fig10}. Note the conjunctions of Mercury-Mars, Sun-Jupiter, while Venus was in Aquarius and the lunar conjunctions with these planets while approaching to the vernal equinox.
 Several years ago Gombrich pointed out \cite{gombrich2} that there exist a letter dated in 1477 sent by Ficino to the teenager Lorenzo di Pierfrancesco, which contained a moral exhortation in the form of an allegorical horoscope to fix his eyes to Venus, the {\it Humanitas}, according to the pedagogical advise of Cicerone to use visual teaching tools. The letter was accompanied by a note addressed to his educators, Giorgio Antonio Vespucci and Naldo Naldi, in order to incite him to memorize the contents since the young Popolano, being an irascible character, did not possess the {\it humanitas} virtue. 

As a conclusive provocation, I would like to point out that the planetary configuration reported on the spring equinox of 1477 did not occur in the years that have been hypothesized for dating Botticelli's {\it Primavera}.

\paragraph{Acknowledgments} \small{The author thanks Mario Gilberto Lattanzi, Sandro Caranzano for the very useful comments which improved the text, especially SC for the suggestion of  {\it Casa di Venere in Conchiglia}; Michele Guido for having pointed out the texts of Marco Paoli and Mirella Levi d'Ancona; Luisa Schiavone for having forwarded the Felice Stoppa's site and the edition of {\it Poeticon Astronomicon} of 1482; Roberto Morbidelli for the botanical confirmation of Laur Nobilis and Myrtle.}


\end{document}